\documentclass{amsart}
\usepackage{amsmath,amssymb,color,cite,hyperref}
\usepackage{rotating}

\newcommand{\p}{{\partial}}

\begin{document}
\title[A Systems Biology Approach to Muscular Dystrophy: A Review]{A Review of Mathematical Models for Muscular Dystrophy: A Systems Biology Approach}
\author{Amanda N. Cameron}
\address{Amanda N. Cameron: University of Georgia, Department of Mathematics, Athens, 30602, USA}
\author{Matthew T. Houston}
\address{Matthew T. Houston: University of Georgia, Department of Mathematics, Athens, 30602, USA,  Middle Georgia State University, Department of Mathematics, Macon, 31206, USA}
\author{Juan B. Gutierrez}
\address{Juan B. Gutierrez: \href{mailto:jgutierr@uga.edu}{jgutierr@uga.edu} University of Georgia, Department of Mathematics and Institute of Bioinformatics, Athens, 30602, USA}
\begin{abstract} 
Muscular dystrophy (MD) describes generalized progressive muscular weakness due to the wasting of muscle fibers. 
%While this is  an umbrella term used to describe a wide range of muscle wasting diseases, the two types that have been most extensively mathematically modeled are Duchenne's (DMD) and Becker's (BMD), with DMD being the most common childhood form of MD. 
The progression of the disease is affected by known immunological and mechanical factors, and possibly other unknown mechanisms. These dynamics have begun to be elucidated in the last two decades.   This article reviews mathematical models of MD that characterize molecular and cellular components implicated in MD progression. A biological background for these processes is also presented.  Molecular effectors that contribute to MD include mitochondrial bioenergetics and genetic factors; both drive cellular metabolism, communication and signaling. These molecular events leave cells vulnerable to mechanical stress which can activate an immunological cascade that weakens cells and surrounding tissues. This review article lays the foundation for a systems biology approach to study MD progression. 
\end{abstract}

\maketitle 

%##############################################
\section{Introduction}
%##############################################
Muscular dystrophy (MD) describes generalized progressive muscular weakness due to the wasting of muscle fibers. While this is  an umbrella term used to describe a wide range of muscle wasting diseases, the two types that have been most extensively mathematically modeled are Duchenne's (DMD) and Becker's (BMD) \cite{DellAcqua2009283,jarrah2014mathematical}, with DMD being the most common childhood form of MD. Males affected by this X-linked recessive disorder have an average life expectancy in the mid-twenties, typically becoming fully wheelchair dependent by their teens \cite{brooke1989duchenne}. MD leaves striated muscle cells with reduced contractile abilities, leading to a wide range of phenotypic expression in patients from fatigue to drooping eyelids.  

Previous research attributes pathogenesis of DMD and BMD to either absent or partial forms of the dystrophin protein \cite{hoffman1987dystrophin}. From a cellular perspective, the basic contractile unit of the muscle is the sarcomere. The dystrophin protein is located between the sarcolemma, the outer membrane of the sarcomere, and outer layer of myofilaments, providing a scaffold for muscular contraction. Weakness typically begins in extremity muscles, propagating in a proximal--distal direction, until ultimately affecting the diaphragmatic muscles responsible for breathing \cite{beggs1991exploring,brooke1989duchenne}. Hypertrophy of cardiac muscle cells is an additional complication associated with both DMD and BMD. The loss of function associated with both typically leads to premature death \cite{burns2015evidence}.

Epidemiological impacts of MD remain difficult to pinpoint definitively. Each subset of MD contains its own range and pattern of pathogenesis and progression. We can further classify subsets of MD in (ordered in decreasing rates of prevalence): dystrophinopathies, laminopathies, dystroglycanopathies, sarcoglycanopathies, and alternative congenital forms \cite{alhamidi2011fukutin,norwood2009prevalence}. In the United States, dystrophinopathies have a prevalence of about 15 cases per 100,000 people \cite{romitti2015prevalence}.  The following list describes the rates of prevalence of MD per 100,000 cases in Northern England \cite{norwood2009prevalence}: 

\begin{enumerate}
	\item With respect to dystrophinopathies that manifest as dystrophin absence or deficiency, DMD and BMD have been most extensively studied. Incidence of dystrophinopathies is about 8.5.
	\item Laminopathies, dystroglycanopathies,and sarcoglycanopathies are typically classified as Limb-Girdle MD (LGMD). The prevalence rate is about 2.
	\item Collagen VI deficiencies that take form in Ullrich Congenital MD (UCMD, also known as Ullrich Scleroatonic MD) have a prevalence rate around 0.13.
	\item Most alternative congenital forms have rates below one. 
\end{enumerate}

The goal of this paper is to offer a survey of quantitative research in MD, as well as identifying opportunities for quantitative research that are yet to be explored. This paper is organized as follows: Section \ref{subsec:Gene} 
%\hyperref[subsec:Gene]{Gene} 
examines gene regulatory networks and their applications in genes correlated with MD, since most types of MD are believed to have genetic origin \cite{hoffman1987dystrophin}. The genes implicated in MD have functions in numerous biological processes, thus obscuring a univocal characterization of MD pathogenesis.  As a consequence, many of these genes affect the normal cell life/death cycle \cite{pauly2012ampk}. 

Section \ref{subsec:Mito} 
%\hyperref[subsec:Mito]{Mitochondrial Models} 
explores mitochondrial and genetic targets that have been identified as major players responsible for the exacerbation of molecular dynamics that contribute to degenerative processes such as apoptosis\cite{tam2013mathematical} and fibrosis \cite{virgilio2015multiscale,desguerre2009endomysial}; we explain in this section the distinction between self-induced and peer-induced apoptosis.  

Section \ref{subsec:Imm} 
%\hyperref[subsec:Imm]{Immunological Models} 
presents the interplay of molecular dynamics that impacts immunological processes. In certain MDs, chronic immune activation leads to a cycle of damage-regeneration that perpetuates disease; as a result, muscle cells are replaced by fibrous and adipose tissue \cite{DellAcqua2009283,lemos2015nilotinib}. Section \ref{subsec:Musc} 
%\hyperref[subsec:Musc]{Muscle Models} 
explores the mechanical and immunological dynamics of muscle cells. Cyclic immunological activation purported by mechanical stress is associated with coexisting restorative and degenerative processes in muscle cells; immunological processes act both as starters and finishers of apoptosis.  CD$8+$ (cytotoxic T-cells) initiate apoptosis in compromised cells, and the cellular remains of apoptosis are disposed of by macrophages.

%##############################################
\section{Molecular Models}\label{sec:Molecular}
%##############################################
%----------------------------------------------
\subsection{Genetic}\label{subsec:Gene}
%----------------------------------------------
Single nucleotide polymorphisms (SNPs) and their associated gene targets have been implicated in MD (Table \ref{SNPs}). Publicly available databases can be used to investigate each SNP's related genes, proteins, and pathways (e.g. SNPedia \cite{cariaso2012snpedia}, OMIM \cite{amberger2015omim}, and 1000 Genome Project \cite{10002015global}). 

%The exploration of gene regulatory networks (GRNs) can inform the influence perturbations in pathogenesis

Gene regulatory networks (GRNs) can provide insight regarding subtleties in disease development; graph theoretic models like Boolean Networks \cite{li2004yeast}, Bayesian graphs \cite{friedman2000using}, and Petri Net \cite{grunwald2008petri} have emerged as important tools for understanding and reconstructing GRNs.  To develop models of GRNs, researchers may employ database like OMIM \cite{amberger2015omim} or computer programs like Snoopy \cite{rohr2010snoopy} that use experimental genetic expression data to reconstruct such networks. GRNs can then be studied to find nodes that have more parent genes, are up-regulated or down-regulated by parent genes, affect more downstream genes, or express more often by different but related diseases like MDs. For time dependent data or complex processes, though, directed graphs may be impossible to construct. Instead continuous models like dynamic Bayesian networks \cite{nachman2004inferring} and temporal Bayesian classifiers \cite{tucker2006temporal} can be constructed. Although continuous models are unable to answer qualitative questions about the GRN, continuous models can estimate quantitative values. Lack of data hampers continuous models ability to explain large complex GRNs. A continuous model of a GRN involving DMD, LGMD2C, and LGMD2E was constructed by Tucker et al. (2006) \cite{tucker2006temporal} using temporal Bayesian classifiers; the model predicts that genes Dlk1, Dusp13, and Casq2 play a part in all three MDs and should be studied further as a possible influence of MD pathogenesis.

\begin{figure}[t]
	\includegraphics[width=12cm]{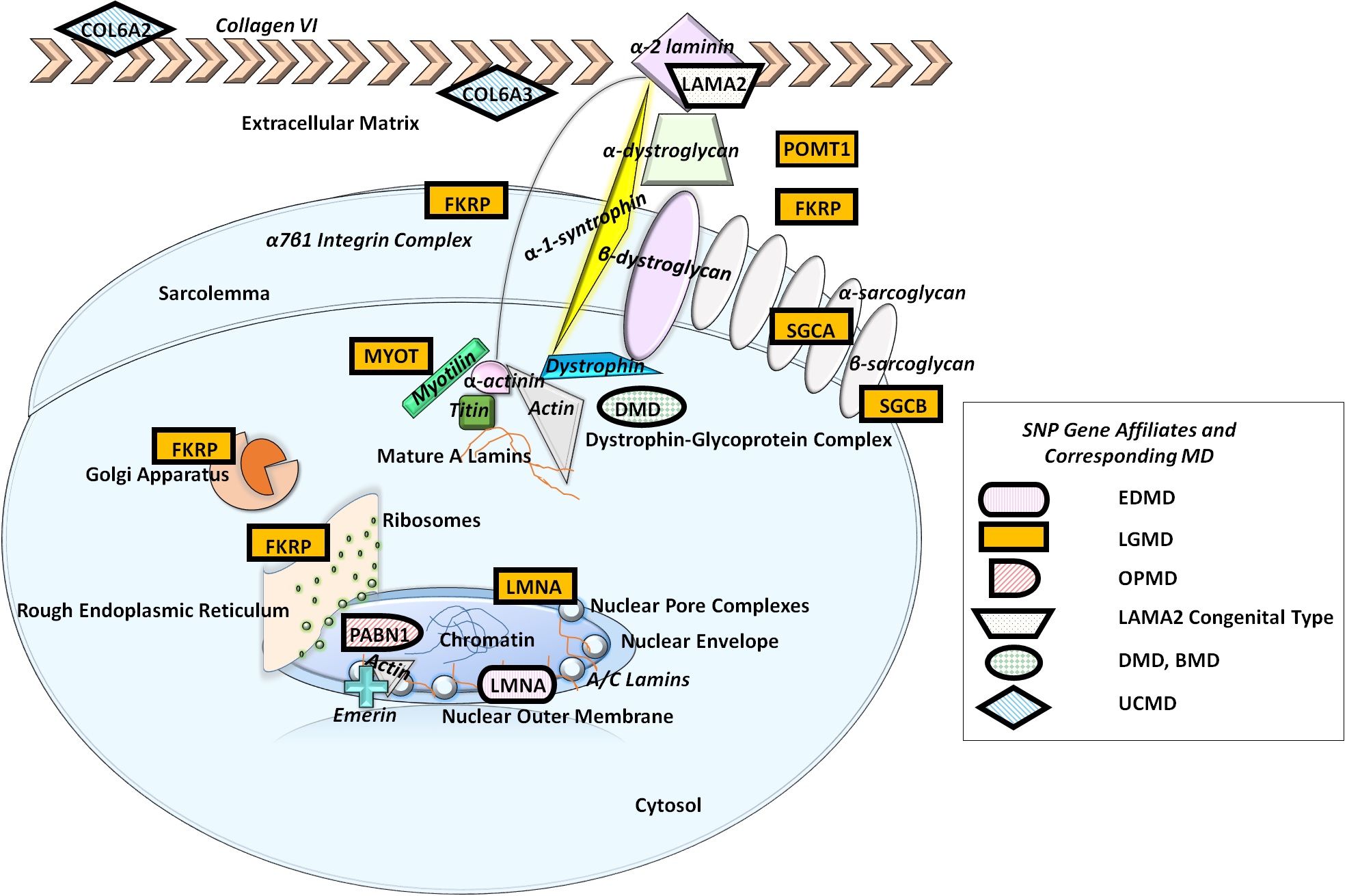}
    \caption{A regional representation of MD type and corresponding gene SNP in a muscle cell.} \label{Cell} 
\end{figure}
\begin{table}[t]
\begin{tabular}{|p{.75 in}p{.75 in}|c||c|c}
\hline
\textbf{SNPs} & & \textbf{Gene Target} & \textbf{MD}\\ 
\hline
rs28928901 & & LMNA & EDMD, LGMD\\
\hline
rs104894466 & & PABPN1 & OPMD\\
\hline
rs28937902 & rs28937903 & FKRP & LGMD \\  
rs28937904 & rs28937900 & & \\
rs28937905 & rs28937901 & & \\
\hline
rs2296949 & & POMT1 & LGMD\\
\hline
rs28933096 & & LAMA2 & LAMA2 Congenital Type MD\\
\hline
rs28937597 & & MYOT & LGMD\\
\hline
rs1800278 & & DMD & DMD, BMD\\
\hline
rs28933693 & rs28933694 & SGCA & LGMD\\ 
\hline
rs28936383 & rs28936384 & SGCB & LGMD\\ 
rs28936385 & rs28936386 & & \\ 
\hline
rs1042917 & rs2839110 & COL6A2 & UCMD\\
\hline
rs1131296 & rs2270669 & COL6A3 & UCMD\\
\hline
\end{tabular}
\caption{SNPs Implicated in MD. SNP IDs correspond to SNPedia.com \cite{cariaso2012snpedia} } 
\label{SNPs} 
\end{table}

In this section we identify the mechanisms associated to the SNPs listed in Table \ref{SNPs}. Since not all these pathways have been modeled quantitatively, this section offers opportunities for systems biology exploration. Fig \ref{Cell} shows the location in the cell of each protein product associating with genes containing SNPs implicated in MD. It illustrates the extracellular and intracellular domains of a muscle cell, relating proteins embedded within and associated with the nucleus, the dystrophin-glycoprotein complex (DGC) and the extracellular matrix (ECM). The nuclear outer membrane (OM) is an integral part of the rough ER (Fig \ref{Cell}). Destabilized lamina release proteins that diffuse along the OM to the ER. 

A and C type lamins (Fig \ref{Cell}) provide structural support within the nuclear membrane and nuclear interior (NI); these are encoded by the gene LMNA (SNPs, Table \ref{SNPs}) \cite{lin1993structural}. Structural deficiencies can cause deficits in gene expression, metabolic signaling within the rough endoplasmic reticulum (ER), and cell cycle control. The ER is a metabolic highway -- with regards to MD, an important target as one featured pathway includes the release of calcium during contraction cycles in muscles  \cite{rolls2002targeting}. 

PABPN1 (Polyadenylate-binding nuclear protein 1) is responsible for the post--transcriptional modification of mRNA and long non-coding RNAs (lncRNAs) within the nucleus, though its cytoplasmic function remains unknown (SNPs, Table \ref{SNPs}). While expressed at high levels in all tissues, mutations are associated with Oculopharyngeal MD (OPMD), pathology is strictly localized to muscles \cite{trollet2010molecular}. One pathway, p53 (Fig \ref{Cell}), is an apoptotic determiner with presence attributed to the expression of the gene PABPN1 \cite{bhattacharjee2012expression}. 

The extent of apoptosis in MD pathogensis is variable among types of MD. In DMD and BMD, apoptosis occurs prior to necrosis, which is initiated after changes in muscle histology. Apoptotic events persist throughout disease progression. Determining deficits which trigger apoptotic events is difficult; MD SNPs have been implicated in apoptotic promotion, though variable among different MD types \cite{tidball1995apoptosis,girgenrath2004inhibition,bhattacharjee2012expression}. Nuclear apoptotic targets also include A lamins as well as LAP2$\alpha$ (Lamina-associated polypeptide 2) and LAP2$\beta$. Depletion of caspase 6, necessary for cleavage of A lamins during apoptosis, is associated with delayed or fully inhibited apoptosis. A lamin degradation is required for timely apoptosis \cite{abu2003involvement, fan2001oligomerization}.

Most common within the nuclear inner membrane (IM) is the protein emerin encoded by the gene EMD; partial or absent forms of emerin are associated with Emery-Dreifuss MD (EDMD).  Mutant A or C type lamins results in the diffusion of improper proteins like emerin in EDMD cells, reacting with ER proteins and hijacking these metabolic pathways \cite{rolls2002targeting,sakaki2001interaction,tsuchiya1999distinct}. Mature A lamins bind to emerin as well as cytoskeletal proteins actin and titin (Fig \ref{Cell}). With regards to EDMD, deficient emerin only leads to skeletal and heart muscle defects \cite{sakaki2001interaction,sullivan1999loss}.

Intracellular FKRP (SNPs, Table \ref{SNPs}) expression has been found in the perinucleus, the ER, Golgi cisernae \cite{brockington2001mutations}; extracellular expression has been found in the sarcolemma and between myofibrils. FKRP coordinates with putative glycosyltransferases (GT) or phosphotransferases (PT) to assist in the packaging of dystroglycan (DG) proteins, encoded by the gene DAG1, as they move to the sarcolemma. $\alpha$-DG hypoglycosylation is the precursor interaction between the actin cytoskeleton and components of the ECM. $\alpha$-DG O-glycosylation is required for extracellular expression of $\alpha$-DG and again in muscles (Fig \ref{Cell}) \cite{alhamidi2011fukutin, brockington2001}. Overexpression of FKRP inhibits maturation of $\alpha$ and $\beta$-DG, post-translation \cite{esapa2002functional}. DG defects are associated with six genes that encode for putative GT or PT; two of which are POMT1 (SNPs, Table \ref{SNPs}) and FKRP (SNPs, Table \ref{SNPs}). POMT1  works with the POMT2 gene product to build a protein complex that enables protein O-mannosyltransferase activity \cite{de2002mutations}. POMT1 mutant genes are associated with limb girdle type 2K with mental retardation as well as congenital MD. Mutations in POMT1 are also associated with an abnormal $\alpha$-DG pattern in the muscle \cite{alhamidi2011fukutin, dinccer2003novel, kim2004pomt1, godfrey2007refining}. 

Laminin-$\alpha$2 is encoded by the LAMA2 (SNPs, Table \ref{SNPs}) gene; it's localized to striated muscle and Schwann cells. Laminin-2, also referred to as merosin, engages with the DGC through binding to $\alpha$-1-syntrophin, a calcium pump \cite{williams2006sarcolemmal}, and the $\alpha$7$\beta$1 integrin complex, a cell surface receptor (Fig \ref{Cell}). Though the overexpression of $\alpha$7$\beta$1 integrin in mice models with deficient levels of laminin-$\alpha$2 does result in a milder MD phenotype, deletion of $\alpha$7$\beta$1 integrin fails to affect mice models of laminin-$\alpha$2 deficient MD. This suggests overlapping functions of proteins within the DGC and ECM \cite{gawlik2015deletion, Gurpur2009999}. 

Myotilin (Myo and Titin Immunoglobulin Domain protein) and laminin-$\alpha$2 form complexes indirectly with dystrophin (Fig \ref{Cell}). Encoded by MYOT (Table \ref{SNPs}), myotilin mitigates sarcomere formation; MYOT defects are associated with LGMD1A \cite{salmikangas2003myotilin, alhamidi2011fukutin}. With regards to laminin-$\alpha$2, $\alpha$-DG binds to $\alpha$-2 laminin in the ECM while $\beta$-DG binds to dystrophin in the transmembrane \cite{ter1998laminin}. 

The dystrophin protein, encoded by DMD (Table \ref{SNPs}) \cite{hoffman1987dystrophin}, acts as a scaffold in a subsarcolemmal space protein complex for muscle cells; dystrophin binds to actin, bridging the extracellular and intracellular domains with the cytoskeleton of the muscle cell (Fig \ref{Cell}) \cite{rybakova2000dystrophin}. Absent or partial forms of dystrophin uncouples the DGC; mechanical stress aside, this also disrupts cellular communication \cite{grady1999role,williams1999differential,judge2006dissecting}.

Sarcoglycan (SG) defects in MD are encoded by the genes SGCA (SNPs, Table \ref{SNPs}) and SGCB (SNPs, Table \ref{SNPs}); these encode for the sarcolemma proteins $\alpha$-SG and  $\beta$-SG that stabilize muscle fiber membranes (Fig \ref{Cell}). Mutant SGCA is rescued by inhibition of proteasome-mediated, ER-associated degradation (ERAD) of mannosidase I \cite{bartoli2008mannosidase}. A compromised sarcoglycan complex is linked with LGMD; histological analysis reports aggregates of peripheral mitochondrial accumulation as well as increased, abnormal, levels of serum creatine kinase \cite{sewry1996abnormalities}. 

Collagen VI is a common ECM protein, with deficiencies associated with COL6A1, COL6A2 (SNPs, Table \ref{SNPs}), and COL6A3 (SNPs, Table \ref{SNPs}) gene defects (Fig \ref{Cell}). UCMD is caused by COL6A2 and COL6A3 mutations. Histological analysis of UCMD muscle reveals an increased number of internal nuclei as well as an observable dystrophic pattern in muscle biopsy samples collected; lax and hyper-flexible joints experienced by patients also result in higher propensities for eccentric contractures (Section \ref{subsec:Musc}
%\hyperref[subsec:Musc]{Muscle Models}
) and myosclerosis. Collagen VI ECM mutations also affect the PI3K-Akt signaling pathway (Figure \ref{Cell}), which has immunological applications (Section \ref{subsec:Imm}
%\hyperref[subsec:Imm]{Immunological Models}
) \cite{baker2005dominant,cheng2011collagen}.

Models for cell motility that integrate cytoskeletal molecular interactions within models for muscular contraction are needed to shed light on genotype to phenotype interactions. With regards to cell cycle control, mechanical models for the assembly and reassembly of the nucleus during mitosis will be important in better understanding the consequences of lamin deficits in MD (Section \ref{subsec:Musc}
%\hyperref[subsec:Musc]{Muscle Models}
).  

With a need to relate discrepancies in phenotypic severity with genotype mutations, graph theoretic models have emerged as important tools.  SNPs associated with MD provide data for building gene regulatory networks. Gene, protein and metabolic regulatory networks require the integration of affiliate pathways and mechanisms within any future mathematical models.
 
%##############################################
\subsection{Mitochondrial}\label{subsec:Mito}
%##############################################
The distinct role of dystrophin in DMD and BMD's pathogenesis and progression remains unknown; despite devastating consequences in its absence, healthy skeletal muscle expresses a mere $0.002\%$ of the dystrophin gene \cite{hoffman1988characterization}. This requires the examination of alternative factors that lead to similar devastating MD phenotypes \cite{hoffman1987conservation, hack1998gamma}.

Poorly regulated mitochondria have been implicated in DMD and BMD \cite{rybakova2000dystrophin}. Mitochondria produce ATP, which is the currency for sarcomere contraction. In a single cell, there can be hundreds of mitochondria; these organelles are responsible for far more than ATP synthesis. Mitochondria synthesize protein encoded in mitochondrial DNA (mtDNA) in addition to dividing independently as needed inside the cell. mtDNA is vulnerable to mutations that perpetuate the mitochondrial cascade, with rates up to ten times higher than nuclear DNA mutation  \cite{kowald1996network,kowald1994towards}. 

Mitochondrial proteins like calmitine also regulate the balance of bound and free calcium in the mitochondrial matrix \cite{lucas1996skeletal}; healthy levels of free calcium initiate oxidative phosphorylation \cite{glancy2013effect}.  Heron's (1995) \cite{lucas1995muscular} research noted defects in genes such as C57 BL 6J dy/dy, attributed to expression of similar DMD phenotypes, such as in mice muscle cell necrosis. Furthermore, both mice models of disease also shared calmitine defiencies. Calmitine is the only mitochondrial protein responsible for binding to calcium, whose expression is contained regionally in the mitochondrial matrix of skeletal fast twitch muscle fibers. These deficits are responsible for higher calcium levels, ultimately activating the mitochondrial cascade attributed to DMD progression. Given DMD's and mitochondrial myopathies' maternal inheritance, Heron's molecular models provide a basis for MD mitochondrial modeling, especially with regards to MD pathogenesis and progression \cite{lucas1995muscular}. 

Mitochondria's shape, number, and energy processing power are constantly in flux; mitochondrial dynamics alternate between fusion or fission. Tam et al. (2013) \cite{tam2013mathematical} produced a stochastic and probabilistic model examining rates of fusion-fission that would optimize mitochondrial function and minimize clonal expansion in neighboring mitochondria. This model classifies fusion-fission events in mitochondria using the following criterion: 
\begin{enumerate}
\item Fusion events feature nucleoid exchange between mitochondria; one is emptied out to the other and marked for degradation.
\item Fission sites appear close to the fusion event, regionally containing original nucleoid distributions from precursor mitochondria.
\item Healthy fission features low levels of exchange of nucleoids, with only mitochondrial matrix contents being mixed. 
\item Larger and longer mitochondria have higher propensities for fission, and smaller ones are more likely to fuse.
\end{enumerate}

The initial conditions in this model feature $a_{R,0},$ the propensity for nucleoids to replicate as well as $a_{fus}$,the propensity for mitochondrial fusion.  All other propensities are computed afterwards, as shown in Table \ref{Prop}.  Protective nuclear retrograde signaling could rescue the mitochondrial cascade through the promotion of mitochondrial nucleoid replication propensity up to sixteen times the basal rate, increasing stochasticity by neutralizing clonal mutant aggregation. Benefits of nuclear retrograde signaling are limited by rates of fusion-fission. Within this simulation, rate of mitochondrial fusion-fission plays a significant role in clonal expansion. Slow exchanges of mtDNA result in homoplasmy, where interventionist retrograde signaling could compound the issue by increasing rate of nucleoid replication. Higher rates of fusion-fission result in a heteroplasmic steady state; increasing levels of mitochondria in cells mix nucleoids faster. These patterns persist regardless of mitochondrial presence in the cell or their replication parameters. A cytoskeletal, cellular model that considers mitochondrial movement independent of fusion-fission, as well as mitochondrial morphology in a differentiated cell context, is needed to further conclude potential therapeutic benefits of retrograde signaling \cite{tam2013mathematical}.

\begin{table}[!t]
%\begin{adjustwidth}{-2in}{-2in}
\begin{tabular}{|p{1.9 in}|c|}
\hline
 Propensity & Propensity equations \\  
\hline
Upregulated propensity for nucleoid replication due to higher ratio of mutant mtDNA & $a_{R,del} = a_{R,0}\left (r_{max} \left (1 - \dfrac{R^m_W}{K^m_{retro} + R^m_W} \right ) + 1\right )$ \\
\hline
Propensity for mitochondrial autophagy  & $a_{D,mito} = k_D N_{Mito}$ \\
\hline
Propensity for mitochondrial fission & $a_{fis,i} = V_{F,\max}\dfrac{(W_i+M_i)^n}{K^n_F + (W_i+M_i)^n}$ \\
\hline
\end{tabular}
\caption{Propensity Equations for Tam et al. (2013) \cite{tam2013mathematical} where $r_{max}+1$ - is the maximum copy number for amplification of mtDNA, $N_{Mito}$ - is the number of mitochondria in a cell at a given time $k_D$ - the rate of autophagy, and $V_{F,max}$ - is the maximum propensity of fission.} \label{Prop}
%\end{adjustwidth}
\end{table}

Byproducts of mitochondrial metabolism include small amounts of electrons that leak from inner membrane complexes and attach themselves to oxygen, forming free radicals called reactive oxidative species (ROS). In a healthy immune response, free radicals such as superoxide and nitric oxide are produced by macrophages for destruction of foreign species. Small amounts of the free radical superoxide produced by the mitochondrion are neutralized by antioxidant enzymes such as superoxide dismutase.  Mutated mtDNA leak more ROS in a degenerative, mitochondrial cascade essentially poisoning vulnerable cells through ROS release \cite{kowald1996network,kowald1994towards}.

\begin{table}[!ht]
%\begin{adjustwidth}{-2in}{-2in}
\begin{tabular}{|c|p{2.5in}|}
\hline
Fusseneger et al. Equations \cite{fussenegger2000mathematical} & Variables  \\
\hline
$\dfrac{d[RL]_T}{dt} = b_{L} - \mu [RL]_T $ & $[RL]$ - Concentration of FAS-FAS ligand complex. \\
$\dfrac{d[R]_T}{dt} = b_R - \mu [R]_T$ & $[R]$ - Concentration of FAS receptor. \\
$\dfrac{d[F]}{dt} = b_F - 2r_F - \mu [F]$ & $[F]$ - Concentration of FADD protein. \\
$\dfrac{d[RL.F_2]_T}{dt} = r_F - \mu [RL.F_2]_T$ & $[RL.F_2]$  - Concentration of FAS-FASL-FADD complex. \\
$\dfrac{d[C_c]}{dt} = r_c - r_{A1} - \mu [C_c]$ & $[C_c]$ - Concentration of cytosolic cytochrome c. \\
$\dfrac{d[Al]_T}{dt} = b_{A1} - \mu [A1]_T$ & $[A1]$ - Concentration of Apaf-1 protein. \\
$\dfrac{d[A1.C_c]}{dt} = r_{A1} - \mu [A1.C_c]$ & $[A1.C_c]$ - Concentration of Apaf-1-cytochrome c complex. \\
$\dfrac{d[C_{8z}]}{dt} = b_{8} - 2r_{8zal} - \mu [C_{8z}]$ & $[C_{8z}]$ - Concentration of procaspase-8. \\
$\dfrac{d[C_{9z}]}{dt} = b_{9} - 2r_{9zal} - \mu [C_{9z}]$ & $[C_{9z}]$ - Concentration of procaspase-9. \\
$\dfrac{d[C_{8a}]}{dt} = 2r_{8zal} - \mu [C_{8a}]$ & $[C_{8a}]$ - Concentration of active caspase-8.  \\
$\dfrac{d[C_{9a}]}{dt} = 2r_{9zal} - \mu [C_{9a}]$ & $[C_{9a}]$ - Concentration of active caspase-9. \\
$\dfrac{d[C_{Ez}]}{dt} = b_{Ez} - \sum \limits_{w=8}^9 {r_{wEa}} - \mu [C_{Ez}]$ & $[C_{Ez}]$ - Concentration of executioner procaspase. \\
$\dfrac{d[C_{Ea}]}{dt} = \sum \limits_{w=8}^9 {r_{wEa}} - \mu [C_{Ea}] - r_{IAP}$ & $[C_{Ea}]$ - Concentration of active executioner caspase. \\
$\dfrac{d[B_2]}{dt} = b_{B_2} - \mu [B_2]$ & $[B_2]$ - Concentration of Bcl-2. \\
$\dfrac{d[B_x]}{dt} = b_{B_x} - \mu [B_x]$ & $[B_2]$ - Concentration of Bcl-X$_L$. \\
$\dfrac{d[I_8]}{dt} = b_{I_8} - \mu [I_8]$ & $[I_8]$ - Concentration of FLIPs. \\
$\dfrac{d[I_9]}{dt} = b_{I_9} - \mu [I_9]$ & $[I_9]$ - Concentration of ARC. \\
\hline 
Huber et al. Equations \cite{huber2010diffusion} & Variables \\
\hline
$\dfrac{\p C_n(x,t)}{\p t} = \dfrac{\p ^2 C_n(x,t)}{\p x^2} + v_n(x,t)$ & $C_n(x,t)$ - concentration of a given protein (n=1,2,..,23) \\
& $v_n(x,t)$ - Chemical reactions given by usual Mass/Kinetic action.\\
\hline
\end{tabular}
\caption{Equations for Mitochondrial Models.} \label{MitoODE}
%\end{adjustwidth}
\end{table}

Intrinsic apoptosis results from a stressed cellular response. Severe ROS damage can result in cellular necrosis whereas the release of cytochrome c to the cytosol from the inner membrane of the mitochondria triggers intrinsic apoptosis. Depolarization in MOMP (mitochondrial outer membrane permeabilization) in stressed cells triggers the MOMP cascade. MOMP stress markers in survival-apoptotic dynamics feature members of the Bcl-2 protein family and BH-3 only proteins.  \cite{fussenegger2000mathematical,rehm2009dynamics}. When triggered by MOMP, mitochondria release cytochrome c and Smac (second mitochondrial-derived activator of caspases) into the cell's cytosol.  By binding with XIAPs (x-linked inhibitors of apoptosis), Smac allows for cytochrome c, Apaf-1, and ATP to combine into apoptosome and cleave procapase-9 forming the initiator caspase-9;  cleavage of procaspase-3 by caspase-9 then forms the executioner caspase-3 which in turn activates executioner caspase-6,7 and creates a positive feedback loop by cleaving more caspase-9.  These executioners finalize the death of the cell \cite{fussenegger2000mathematical,rehm2009dynamics,huber2010diffusion}. Apoptosis will not occur if threshold levels of effector capases are not reached \cite{fussenegger2000mathematical,rehm2009dynamics, huber2010diffusion}.

Extrinsic apoptosis occurs when an extracellular self-destruct order is given. Stressed cells signal macrophages to engage with apoptotic cells through phagocytosis as a protective mechanism. Extracellular death ligands act as messagers of these orders by binding with FAS (CD95) death receptors.  Subsequently, FAS receptors cluster allowing the binding with FADD (FAS-associated death domain).  Recruited by FADD, multiple procapase-8 compile and mutually cleave forming capase-8. This new protein cleaves pro-apoptotic, BH3-only protein Bid forming tBid (Truncated Bid). These interactions lead up to MOMP activation and subsequent cascade; the MOMP activation pathway bridges extrinsic and intrinsic apoptosis. Multidomain proteins are activated by apoptotic tBid activation, which can be inhibited due to protective Bcl-2 proteins \cite{fussenegger2000mathematical,rehm2009dynamics}.

Fussenegger et al. (2000) \cite{fussenegger2000mathematical} proposed a model simulating apoptosis to study caspase activation and inhibition (See Table \ref{MitoODE}). The model confirms experimental observations that Bcl-2 above a critical level effectively inhibits procaspase-9 activation but fails to adequately inhibit procaspase-8 activation, and suppression of FADD's binding to FAS/FASL complex blocks caspase-8 activation but has little effect on caspase-9 activation. The model assumes isotropic reactions with a well mixed single domain and omits proteins including Bid/tBid, reactions like caspase-8 cleaving of Bid, and bundles executioner caspases 3,6,7 into a single variable.  Furthermore, intrinsic and extrinsic apoptosis were not distinguished.  Since Fussenegger et al., several models have been proffered to redress omissions.

Albeck et al. (2008) \cite{albeck2008modeling,albeck2008quantitative} introduced a  model concentrating on the extrinsic apoptosis death switch as well as MOMP interactions.  The model represents both cytosol and mitochondria as two separate domains interacting after MOMP with parameters trained by live-cell imaging of HeLa cells. Similar to Fussenegger et al., Albeck et al. bundles many proteins with similar properties -- such as caspase-8 and -10 are represented as a single variable C8 -- to simplify the model and all reactions are isotropic.  However, unlike the previous model, Albeck et al. incorporates a time delay mechanism to compensate for the delay of death ligand reception to MOMP as oppose to the quick death of the cell post-MOMP.  Intriguingly, western blot fails to show enough XIAP pre-MOMP to properly inhibit caspase-3.  Albeck et al. concluded that another protein/reaction must exist to account for this discrepancy.  The model did confirm that MOMP occurs after proapoptotic Bcl-2 proteins reach a certain level depended on the physiological state of the cell. An alternate stable state -- partial cell death -- is predicted by the model.

To eliminate the isotropic assumption, Huber et al. (2010) \cite{huber2010diffusion} combined previous models with one-dimensional diffusion PDEs.  The typical mass reactions and kinetics are extended by a PDE (Table \ref{MitoODE}) where $v_n(x,t)$ is the chemical reactions.  Although their goal was to investigate anistropic reactions in MOMP, the reaction-diffusion equations remain applicable to wider studies. 

%cite
%Mitochondrion manage the cellular death process; these organelles determine whether the cell dies through necrosis or apoptosis. Apoptosis is a protective metabolic process as opposed to necrosis. Apoptosis signals macrophages to engulf the cell through phagocytosis; with necrosis, the cell membranes are lysed with toxic contents of the cells freely floating. Severe ROS damage can result in necrosis of the cell; the release of cytochrome c triggers intrinsic apoptosis

%##############################################
\section{Cellular Models}\label{sec:Cellular}
%##############################################
%----------------------------------------------
\subsection{Immunology}\label{subsec:Imm}
%----------------------------------------------

Major players in DMD immune response are macrophages and cytotoxic T cells. Macrophages are innate immune system members that lie in wait within mucous membranes, activated by the presence of foreign species in the body. Activated M1 macrophages are phagocytes that engulf these species as a threat to tissues and organs; they initiate the inflammatory response responsible for further recruitment as needed. The adaptive immune system is activated post-phagocytosis when macrophages present antigens to corresponding helper T cells. Helper T cells coordinate an immune response to recruit members such as Cytotoxic T cells. These initiate extrinsic apoptosis (Section \ref{subsec:Mito}
%\hyperref[subsec:Mito]{Mitochondrial Models}
), targeting damaged cells that display Class I MHC markers which indicate a threat to healthy peers. Cytotoxic T cells inject granzymes that destroy the targeted cell. Neutralization of the threat in a healthy immune response occurs at a threshold; M2 macrophages promote tissue repair due to damage caused by the inflammatory response \cite{arnold2007inflammatory,mills2000m,st1994differential}.

Two models have been proposed using a predator-prey system to mathematically model damage induced immune response in DMD, employing a log-normal distribution:
\begin{equation} \label{lognorm}
\alpha(t) = \dfrac{h}{t\sigma \sqrt{2\pi}} e^{\stackrel{-\underline{(\ln{(t)} - m)^2}}{2\sigma^2}}
\end{equation} 
to analyze initial damage. Both models utilize concentrations of the previously listed immune response helpers in addition to concentrations of healthy, damaged, and regenerating muscle cells.  Dell'Acqua and Castiglione (2009) \cite{DellAcqua2009283} is generated by five ODEs in addition to a conservation law (Equations \ref{eq:DellAqua}) describing the immune response of DMD in the mice model. They used COPASI's optimization methods on experimental \textit{mdx} mice data from Hoops et al. (2006) \cite{hoops2006copasi} to find the best fit parameters. The set of equations used is
\begin{align} \label{eq:DellAqua}
    \dfrac{dM}{dt} & = b_m + k_1MD - d_MM, \nonumber\\
    \dfrac{dH}{dt} & = b_H + k_2MD - d_HH, \nonumber \\
    \dfrac{dC}{dt} & = k_2HD - d_CC, \nonumber \\
    \dfrac{dN}{dt} & = k_4R - k_5CN - \alpha(t) N, \nonumber \\
    \dfrac{dD}{dt} & = k_5CN - k_6MD - d_DD + \alpha(t) N, \nonumber \\
    100 & = N+D+R, 
\end{align}
where  $M = $ concentration of macrophages; $H = $ concentration of CD4$+$; $C = $ concentration of CD8$+$;  $N = $ percentage of normal muscle fibers; $D = $ percentage of damaged muscle fibers; $R = $ percentage of regenerating muscle fibers. 

Jarrah et al. (2014) \cite{jarrah2014mathematical} (Equations \ref{eq:Jarrah}) refines Dell'Acqua-Castiglione's model by adding an additional ODE which allows the conservation law to be implicit.  The parameters of this model were derived from recent experimental data of \textit{mdx} mice. Both models assume that the missing dystrophin in the muscle causes damaged muscle cells to initiate the immune response which contributes to their own damage until eventual apoptosis. The set of equations used is
\begin{align} \label{eq:Jarrah}
	\dfrac{dM}{dt} & = b_m + k_1MD - d_MM,  \nonumber \\
	\dfrac{dH}{dt} & = b_H + k_2MD - d_HH,  \nonumber \\
	\dfrac{dC}{dt} & = b_C + k_2HD - d_CC,  \nonumber \\
	\dfrac{dN}{dt} & = k_4R - k_5CN - \alpha(t) N,  \nonumber \\
	\dfrac{dD}{dt} & = k_5CN - k_6MD - d_DD + \alpha(t) N, \nonumber  \\
	\dfrac{dR}{dt} & = k_6MD + d_D D - k_4 R,
\end{align}
and the definition of variables is the same as Equation \ref{eq:DellAqua}. Initial conditions have levels of cytotoxic T cell levels at 0;  this changes when the impulse damage represented by equation \ref{lognorm} sets the system into motion. When $h=0,$ the impulse damage is negated and the system remains in a stable state; allowing $h>0,$ the model ensures that T helper cells draw cytotoxic T cells to the damaged region. Damage caused by the immune system reaches a peak in weeks four through eight until the presence of the players wanes. By week twelve, the decreased presence of macrophages, CD4+ and CD8+ T cells (week fourteen) results in diminished levels of degeneration and restoration.

Both models display regions of bistability. Depending on the initial damaged caused and $M_0,$ the system collapses to healthy muscle stability or approaches a stability with heterogeneous mixtures of healthy and damaged muscle. This suggests that immune response to muscle damage could be a major contributor to DMD's pathophysiology \cite{DellAcqua2009283,jarrah2014mathematical} which has been shown experimentally \cite{spencer2001helper,wehling2001nitric}.

%This conclusion forms the backbone to several therapeutic options in DMD.  

Both Dell'Acqua-Castiglione and Jarrah et al. models are a broad overview of the processes involved and leave out several proteins, signal pathways, and reactions.  Cytotoxic T cells act as the only non-impulse perpetrator of muscle damage in both models yet fail to elucidate on the mechanics on which cytotoxic T cells cause the damage. Extrinsic apoptosis caused by cytotoxic T cells has been proposed as a possible path \cite{spencer1997myonuclear}.

These models indicate that the strength of the immune response and maintenance of the positive feedback system relies upon moving past these threshold points to enter another stability state. Driving the system into these recovery regions could prove to be a potential therapeutic target for redressing the role of the inflammatory response. 
 
%While a more robust model is needed, their model indicates a period of bistability, a threshold past where damaged muscle cells will either completely recover or coexist in a region of muscle degeneration or regeneration. While there are limitations with a linear model, their non-autonomous system found similar points of instability and dynamics.

%----------------------------------------------
\subsection{Muscle Models}
\label{subsec:Musc}
%----------------------------------------------

One muscle cell contains thousands of sarcomeres interlaced and primed for contraction. A single sarcomere is made up of parallel thin and thick filaments called actin and myosin; these entwined filaments pull on each other during contraction \cite{huxley1974muscular,hall2015guyton}. MD leaves muscles vulnerable to unhealthy contraction which exacerbates damage in a cycle that proves catastrophic to muscle cells even in healthy populations. For diseased muscle, intense activity takes everyday, hourly forms perpetuating MD progression \cite{childers2002eccentric,duclos1998progressive}. 

\begin{figure}[ht]
	\includegraphics[width=12cm]{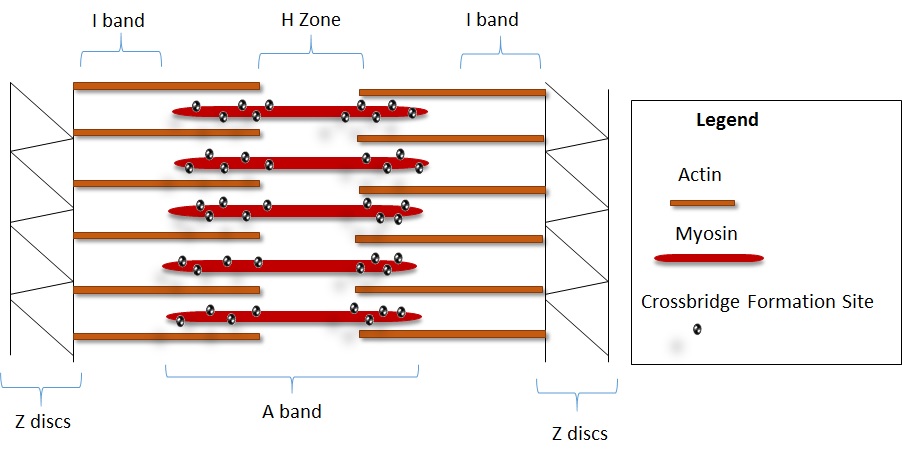}
    \caption{Sarcomere Regional Classification.} \label{sarcomere} 
\end{figure}

We can classify the regions of the sarcomere to better understand contraction mechanics. The A band is made up of thick myosin filaments; throughout contraction its length remains constant centrally localized to the H zone. Actin filaments are laced with the protein tropomyosin. At rest, tropomyosin covers the hinges that catch on actin. Alternatively, the I band is composed of thin actin filaments that alter its length between myosin filament pairs. Structures called Z discs fetter actin filaments at their opposite ends \cite{huxley1974muscular,hall2015guyton}. In S1 regions, myosin edges are hinged and highly flexible, catching on actin and releasing in an indefinite binding cycle (Figure \ref{sarcomere}). Upon their release, myosin filaments perform a power stroke catalyzed by the hydrolysis of ATP. ATP is the means for crossbrige formation; it is the release of phosphate during ATP hydrolysis that contracts the S1 region \cite{huxley1974muscular,hall2015guyton}. Calcium provides the means for binding whereas ATP drives contraction. Upon the release of calcium, the protein troponin pushes the tropomyosin to expose binding sites for actin. Upon exposure and a threshold level of ATP expression, the contraction cycle begins \cite{lehman1994ca2+}.

Huxley (1957) \cite{huxley1957muscle} proposed an early mathematical description of this power stroke process. Distance from binding site to crossbridge is taken as the independent variable while function $n(x,t)$ is the probability of a crossbridge being bound at position $x$ at time $t$.  This yields a conservation law of:
\begin{equation}
\dfrac{\p n}{\p t} - v(t) \dfrac{\p n}{\p x} = (1-n)f(x) - ng(x).
\end{equation}
The rate of energy release, $\phi$, by ATP is:
\begin{equation}
\phi = \rho \epsilon \int_{-\infty}^{\infty} {(1-n(x,t))f(x)\ dx} 
\end{equation}
where $\rho$ is the number of crossbridges at $x$, and $\epsilon$ is the energy released by a single crossbridge. This model captures the key physiological properties of crossbridges but fails to illuminate the biochemical interactions; Pate (1989) \cite{pate1989model} proposed a model based on the Huxley model rectifying the biochemical interaction issue.

Nuclear lamina deficiencies are linked with cytoskeletal defects in muscle cells \cite{sullivan1999loss,sewry2001skeletal}. Destabilized DGC due to cytoskeletal compromise within the plasma membrane leads to the aggregation of mechanical stress in contracting muscle cells \cite{lammerding2004lamin}. Histology in lamina deficient diseases commonly presents with maligned and internalized nuclei in muscle fibers \cite{sullivan1999loss,sewry2001skeletal}. Chronically strained A/C deficient lamins result in cells experience higher rates of apoptosis \cite{rao1996lamin, lammerding2004lamin}. Intriguingly, regenerating muscle fibers also share high levels of internalized nuclei- the distinction between pathology and regeneration is unclear presently \cite{sewry2001skeletal}.

Lammerding et al. (2004) \cite{lammerding2004lamin} modeled nuclear misalignment and cytoskeletal stress in lamin A/C-deficient mouse embryo fibroblasts. A/C lamins form compartments for splicing factors as well as RNA polymerase II transcription. Inhibition of A/C laminis suppresses RNA polymerase II-dependent transcription in mammalian cells, while its as a scaffold in nuclear compartments remain unknown. A sinusoidal force with amplitude 0.6 nanonewtons (nN) and with frequency 1 Hz, offset 0.6 nN, was applied through a magnetic trap. Cylindrical coordinates $(r,\theta)$ were used to measure bead displacement with the magnetic bead at the origin and $\theta=0$ for the force direction. The equation represents the induced strain field described by the analytic cell mechanics model proposed by Bausch et al. (1998) \cite{bausch1998local} $u_r$ is the radial component of the induced bead displacement as a function of the applied force, F, cell stiffness $\mu*$ the characteristic cut of radius $\kappa^{-1}$, the distance from the magnetic bead center $r$, and the polar angle $\theta$. 
\begin{equation} \label{Bessel}
u_r(r) = \dfrac{F}{2\pi \mu^*} \left ({cos(\theta)}\dfrac{3(1-\sigma)}{4}K_0(\kappa_1r)-\dfrac{K_1(\kappa r)}{\kappa r}+\sqrt{\dfrac{(1-\sigma)}{2}}\dfrac{K_1(\kappa_1r)}{\kappa r} \right )
\end{equation} 
$K_0$ and $K_1$ are modified Bessel functions of respective order 0 and 1 with 
\begin{equation} \label{Parameter}
K_1(r) = \kappa \left (\dfrac{(1-\sigma)}{2} \right )^{1/2}
\end{equation} 
Fitting bead displacement data to equation \ref{Bessel}, parameters $\mu_*$ and $\kappa$ can be obtained letting $\sigma = 0.5$ and metallic bead contact radius of 2 $\mu m.$

In this model, the cytoskeleton was exposed to the same biaxial strain as the membrane; for each cell type, nuclear deformation increased approximately linearly with applied membrane strain. Both lamin A/C deficient cells result in decreased nuclear stiffness and altered nuclear mechanics. Under resting conditions, they found that the integrity of the nuclear envelope was maintained in A/C deficient cells. However, under pressure, the control cells maintained nuclei integrity far more than lamin deficient cells, though both could be ruptured. Increased vulnerability of A/C deficient cells to mechanical stress comparatively was also concluded, with an increase in both necrotic and apoptic cell fraction\cite{lammerding2004lamin}.
Necrosis mediated through nuclear rupture is not wholly attributed to vulnerability to mechanical stimulation in this model; only about 3-5\% of A/C deficient nuclei ruptured. 

There are limitations with \textit{mdx} mice models to study eccentric contraction in limb muscles, as mice models of the disease resemble degeneration patterns most closely with human diaphragmatic muscles. Modeling gait progression as it relates to degeneration in human dystrophic muscle is essential for quantifying damage due to the cyclic activation of the immune system due to eccentric contractions \cite{pauly2012ampk}. 

Levels of healthy isometric force production can be maintained under the strain of intense activity. Rest commenced with the buffering of calcium levels in the cytosol reduces mechanical stress to a point; inappropriate high or low levels of activity leave muscles vulnerable to damage. Furthermore, both types of activity are associated with increased levels of cytosolic calcium additionally linked with fiber damage and apoptosis \cite{pauly2012ampk}. 

A.V. Hill (1938) \cite{hill1938heat} created a Force-Velocity relationship model to understand isotonic and isometric muscle contractions; his work provides a basis for multibody, dynamic musculoskeletal modeling and simulation.  Hill's model relates rate of muscle contraction (shortening length), $v$, to the load $p$ by: 

\begin{equation} \label{Hill}
(p+a)v = b(p_0-p)
\end{equation}
for constants $a$ and $b$ given by experimental data and $p_0$ is the isometric force. Although Jewell and Wilkie (1958) \cite{jewell1958analysis} demonstrated that this model lost accuracy when exposed to sudden changes to muscle length, Hill's model remains pivotal and is integrated in many other models. 

Van der Linden et al. (1998) attempted whole tissue interaction simulations examining aponeurosis/muscle under stress, limited by computational power, they were restricted to 2D and simplified 3D models \cite{van1998revised}\cite{van1998modelling}. Johansson et al. (2000) \cite{johansson2000finite} improved upon van Leeuwan-Kier's (1997) \cite{van1997functional} model of squid tentacles by separately modeling active muscle attributes with Hill's force-velocity model and describing passive elements as a hyperelastic material. Unfortunately since they used parameter values based on van Leeuwan-Kier's research and ANSYS, Johansson et al. failed to significantly improve upon the predictions made by van Leeuwan-Kier. Yucesoy et al. (2002) \cite{yucesoy2002three} introduced a similar model expanding on Van der Linden's work.  Like Johansson et al., Yucesoy et al. separated muscle into the extracellular matrix (passive) and myofiber (active); from that they created a linked fiber-matrix mesh that fuses a passive element and an active element.  This ``two domain'' approach allowed a glimpse into the interaction of muscle fibers and the extracellular matrix. 

In an effort to understand the effects of geometries on muscle tissue, Sharafi and Blemker (2010) \cite{sharafi2010micromechanical} devised a model where actual rabbit muscle biopsies were estimated with linear functions.  They captured both the geometries of fibers (microscopic level) as well as the fascicles (macroscopic).  Since they were modeling muscle stress, Sharafi-Blemker only modeled the passive elements of muscle which were described as a hyperelastic, nearly incompressible material. In opposition to assumptions made by earlier modelists, Sharafi-Blemker discovered that fascicle display anisotropic characteristics. Incorporating these geometries, Virgilio et al. (2015) \cite{virgilio2015multiscale} created a simulation to study the effects of various disease pathologies including MD.  The simulation uses an agent based system to create fascicle geometry with the addition of fibrosis and fatty tissue infiltration followed by the use of micromechanical model described by Sharafi and Blemker. This model independently verified the results that fibrosis aggravates the symptoms of DMD using only \textit{in silico} methods.

MD patients can experience higher-degree of ankle plantarflexion due to contractures; ``toe-walking'' makes plantar flexor muscles such as the gastrocnemius, soleus and peroneus therapeutic targets.  Although the model concerns with long term use of high heels, Z\"{o}llner's model (2015) \cite{zollner2015high} could be use to describe toe-walking in MD. Due to removal of sarcomere, fascicle shortens with long term toe walking causing less than optimal motion range and increased stress. Healing of damaged muscle fibers occurs upon the fusing of differentiated myocytes alongside the damaged fibers \cite{pauly2012ampk}.

%a single instance of this leads to immediate loss of performance lasting up to 30 days in healthy muscles.  Redressing the damage caused by eccentric muscle contraction requires a 30 day period of simultaneous degenerative and restorative processes. In healthy muscles,Strong forces distributed axially to actin and myosin crossbridges can cause eccentric contraction responsible for muscle damage. DMD patients are more susceptible to eccentric contractions, resulting in increased activation of the immune system (\ref{subsec:Imm}). Eccentric muscle contraction is catastrophic. Healing of these fibers occurs upon the fusing of proliferated and differentiated satellite cells alongside the damaged fibers \cite{pauly2012ampk}. %A univocal characterization of MD requires previous molecular 

%##############################################
\section{Areas of Future Quantitative Research} 
\label{sec:Quant}
%##############################################

Characterizing pathology and pathogenesis of MD requires further study; future models could accelerate therapeutic discovery by testing potential pathways \textit{in silico} as well as detecting new therapeutic pathways. There are areas for computational research that have not yet been explored mathematically. These pathways may be worthwhile exploring to better understand MD disease development and providing treatments that may delay onset or progression.

The creation of rAAV/AAV1 (Recombinant Adeno-associated virus, Adeno-associated virus 1) delivery system and CRISPR/Cas system indicates the possibility of a genetic cure for MD \cite{pozsgai2014gp,long2014prevention}. Heller et al. (2015) \cite{heller2015human} overexpressed the human $\alpha$7 integrin Gene, ITGA7, using the  AAV1 delivery system in \textit{mdx} mice. ITGA7 is a skeletal muscle laminin receptor (Figure \ref{Cell}) whose overexpression does not cause an immune response in \textit{mdx} mice. Protective benefits of the DGC were restored with  ITGA7 overexpression; lifespans were also prolonged. Xu et al. (2015) \cite{xu2015crispr} used CRIPSR/Cas9 to remove mutated exon 23 with the dystrophin genomic region to restore dystrophin expression in the DGC of \textit{mdx} mice.  Clinical trials are currently taking place for LGMD2D \cite{mendell2010sustained}.  Mendell et al. (2012) \cite{mendell2012gene} wrote a review outlining future work in gene therapy. Both systems require a relatively small number of injection sites. Population dispersion models and GRNs could help in the development and effective administration of both rAAV/AAV1 and CRISPR/Cas systems. Population dispersion and diffusion models could be used to predict the outcome from a series of injections to the spread of the corrected genes throughout the body.  Potentially, these models could indicate the most effective injection sites. Beyond the correction of defective genes, both systems could target genes that promote muscle growth and regeneration \cite{rodino2009inhibition}.  GRNs may help in developing new therapies by finding pathways that both system could target and thereby accelerate the body's muscle regeneration. These types of treatments could be applicable to both MD patients and in sports medicine. 

Myostatin -- a TGF-$\beta$ (Section \ref{subsec:Imm}
%\hyperref[subsec:Imm]{Immunological Models}
) protein -- has long been recognized as a possible therapeutic option for MD. Early murine testing for several MD phenotypes produced mixed results.  For DMD (\textit{mdx} mice) and LGMD2F (scgd$^{-/-}$ mouse), Parsons et al. (2006) \cite{parsons2006age} concluded positive results for mice treated early in development before widespread necrosis occurred. Treating sgcg$^{-/-}$ mice (modeling LGMD2C) resulted in positive muscle physiology including increase fiber size, muscle mass, and grip force in addition to reduce frequency of apoptosis; however, muscle histology remained unfazed signifying lack of pathology change \cite{bogdanovich2008myostatin}.  A possible solution proposed by Rodino-Klapac (2009) used AAV1 to genetically edit Follistatin (the major myostatin inhibitor).  A few experiments show that the new treatment has few side effects and shows similar improvements as other myostatin inhibitors in LGMD2A (Calpainopathy). Future research is needed to show if myostatin inhibition is a means to maintain pathophysiology in MD patients \cite{rodino2009inhibition}. GRNs and physiological muscle models could be used to understand effective usage of myostatin.  GRNs could discover new inhibitors and enzyme activators of myostatin; this may allow targeted genetic editing to regulate myostatin similar to Rodino-Klapac (2009) \cite{rodino2009inhibition}. GRNs might also explain why these methods will work with some types of MD but fail with other types. Physiological models may demonstrate the increase in fiber size, muscle mass, and grip force due to myostatin therapies.  

A plethora of models could be used to describe cellular, molecular, and pharmaceutical interactions of the immune system. Precise molecular, mathematical models bridging arginine metabolism with oscillations in macrophage phenotypic expression could be used to model nitric oxide mediated cytotoxicity as well as fibrosis during satellite cell proliferation. Mechanical models for macrophage infiltration and molecular models for macrophage phenotype oscillation could also be useful to better characterize chronic immune system activation due to structural defects. Integration into musculoskeletal simulation may be useful to model the immunological role in MD pathogenesis and progression. Agent based models could be used to imitate immune cells.  

MD disease progression also results in alterations to pathophysiology such as gait and muscle atrophy. Noninvasive studies with patients, especially children, could be critical to create a staged model for gait devolution and morphology. Quantifying degrees of eccentric contraction using musculoskeletal simulation could possibly explain selective degeneration in DMD \cite{hu2015musculoskeletal}.

%##############################################
\section{Discussion}
%##############################################

%With new developments in computational power, a growing amount of research has concentrated on mathematical models for understand biological systems.  The hope is that by creating effective \textit{in vitro} and \textit{in silico} models we will be able to study pathology/progression of a disease and test new theories/medicine without need of a living subject \cite{hodgson2015modelling}.  Using \textit{in vitro} models allow for tactile control over any experiment.  Over the last few decades, collagen hydrogel with muscle derived cells (CHMDCs) have promised to revolutionize \textit{in vitro} experiments and tissue engineering.  For CHMDCs to reach the envisioned use, verification by use of mathematical simulations are needed.  Recently while examining shape and design, Hodgson (2015) \cite{hodgson2015modelling} used a combination finite elements and agent based analysis to illustrate the lines of principle strain and cell migration in CHMDCs confirming earlier work by Smith et al. (2012) \cite{smith2012characterization} which used the \textit{in vitro} model. Future studies are needed to examine muscles such as proximal ones which take on larger loads for longer, especially in DMD and BMD muscles left more fragile. As MD is a rare disorder, the use of mathematical models to understand the underlining systems will allow researches to more efficiently wield the limited subject pool.  
With new developments in computational power and data availability, a growing amount of research is using a systems biology approach to understand pathogenesis and progression of disease. Effective and integrated \textit{in vitro} and \textit{in silico} models could inform biological phenomena, even without the need of a living subject.  For instance, over the last few decades, collagen hydrogel with muscle derived cells (CHMDCs) have promised to revolutionize \textit{in vitro} experiments and tissue engineering.  For CHMDCs to reach the envisioned use, verification by use of mathematical simulations are needed.  Recently while examining shape and design, Hodgson (2015) \cite{hodgson2015modelling} used a combination finite elements and agent based analysis to illustrate the lines of principle strain and cell migration in CHMDCs confirming earlier \textit{in vitro} work by Smith et al. (2012) \cite{smith2012characterization}. As MD is a rare disorder, the use of mathematical models could help elucidate the underlining mechanisms of the disease that might not be easily detectable given the limited subject pool.  

%Although genetic studies have implicated genes as the cause of many types of MD, relatively little is know about about common pathways between these genes that may affect pathogenesis and create similar phenotypes; this necessitates the use of mathematical models describing GRNs (Section \hyperref[subsec:Gene]{Gene}).  Common genes like Dik1, Dusp13, and Casq2 \cite{tucker2006temporal} and there downstream pathways provide future prospects for therapeutic intervention. Collecting genetic data for other types of MD will provide a stronger correlation of pathogenesis for any common genes discovered.  
Although genetic studies have implicated genes as the cause of many types of MD, relatively little is know about about common pathways between these genes that may affect pathogenesis and create similar phenotypes; this necessitates the use of mathematical models describing GRNs (Section \ref{subsec:Gene}
%\hyperref[subsec:Gene]{Gene}
).  Common genes like Dik1, Dusp13, and Casq2 \cite{tucker2006temporal} and there downstream pathways provide future prospects for therapeutic intervention. %Collecting genetic data for other types of MD will provide a stronger correlation of pathogenesis for any common genes discovered. 

Mitochondrial and immunological mechanisms further MD progression.  Mathematical models of apoptosis and mitochondrial fission/fusion help to understand intracellular processes that directly affect cellular vitality and death. Tam et al. \cite{tam2013mathematical}'s mitochondrial fission/fusion model related mitochondrial health with nuclear mechanisms to rescue mutated mtDNA; unhealthy mitochondria can trigger intrinsic apoptosis with the release of cytochrome c.  Fusseneger et al. \cite{fussenegger2000mathematical} and Albeck et al. \cite{albeck2008modeling,albeck2008quantitative} furthered the study of apoptotic mechanics by creating models to simulate the processes; both models agree with already published results.  Extrinsic apoptosis can be signaled by immune cells.  Immunological mathematical models display a larger view of cellular interactions that bridge the gap of molecular actions and tissue level muscle damage. Damage caused by stress and weakened cellular structure is repaired and debris removed by extrinsic apoptosis and phagocytosis.  Dell'Acqua and Castiglione \cite{DellAcqua2009283} and Jarrah et al. \cite{jarrah2014mathematical} created models describing the cellular/tissue interactions of a few immune cells.
	
%Vital to understanding the long term development of most types of MD are mathematical models of of muscles under contractile stress. Partial differential equations and agent based models of anisotropic strain from contraction display where cellular rupture and immune response will likely occur.  Physical therapeutic and pharmaceutical interventions can be targeted to such areas of high stress to stymie MD progression. 
Mathematical models of muscles under contractile stress are essential  to understanding the long term development of most types of MD. Partial differential equations and agent based models of anisotropic strain from contraction display where cellular rupture and immune response will likely occur.  Physical therapeutic and pharmaceutical interventions can be targeted to such areas of high stress to stymie MD progression. Expanding cellular models of the immune response and combining with molecular signals could create a more comprehensive view of muscle tissue regeneration and damage caused by chronic inflammation.  Future research could incorporate multiple levels of models into a unified simulation to give a whole view of the progression of MD. 

Outside of MD, immunological, mitochondrial, and genetic components covered in this review play a role in diseases with higher rates of prevalence such as Alzheimer's \cite{Pieczenik200784}, Parkinson's \cite{Winklhofer201029}, cancer \cite{Pieczenik200784} and ALS \cite{Cozzolino201254}. Aforementioned diseases have strong ties to mitochondrial dysfunction and inflammatory responses that similarly exacerbate disease pathogenesis and subsequent progression. Even the (relatively) benign aging process is associated with mitochondrial dysfunction \cite{kowald1996network}. All enact a financial and emotional cost for those affected and their families. Future research that sheds light on MD disease dynamics, which is likely to  occur through mathematical modeling, will provide the means to engage and perhaps ultimately bypass biological systems coordinating together to exacerbate degeneration. 

%##############################################
\section*{Conflict of Interest(s)}
%##############################################
None of the authors have a financial conflict of interest. A.C. was diagnosed with Progressive Mitochondrial Myopathy in 2010 through a muscle biopsy and differential diagnosis. She worked for the Foundation for Mitochondrial Medicine since 2013 as an intern and now is a volunteer as needed. M.H. was differential diagnosed (and genetically confirmed recently) with Limb-Girdle Muscular Dystrophy in 2004.

%##############################################
\section*{Author Contributions}
%##############################################
J.G. conceived of the paper and advised A.C. and M.T. on the organization of the manuscript. A.C. conceived of the mitochondrial subsection and created all figures and table 1. M.T. conceived of the muscle subsection, and created tables 2-4. Both A.C. and M.T. contributed to the genetic, mitochondrial, immune, muscle, areas of future quantitative research and discussion sections.  All authors reviewed the manuscript.

\bibliographystyle{abbrv}
\bibliography{library}

\begin{thebibliography}{100}

\bibitem{abu2003involvement}
A.~Abu-Baker, C.~Messaed, J.~Laganiere, C.~Gaspar, B.~Brais, and G.~A. Rouleau.
\newblock Involvement of the ubiquitin-proteasome pathway and molecular
  chaperones in oculopharyngeal muscular dystrophy.
\newblock {\em Human molecular genetics}, 12(20):2609--2623, 2003.

\bibitem{albeck2008quantitative}
J.~G. Albeck, J.~M. Burke, B.~B. Aldridge, M.~Zhang, D.~A. Lauffenburger, and
  P.~K. Sorger.
\newblock Quantitative analysis of pathways controlling extrinsic apoptosis in
  single cells.
\newblock {\em Molecular cell}, 30(1):11--25, 2008.

\bibitem{albeck2008modeling}
J.~G. Albeck, J.~M. Burke, S.~L. Spencer, D.~A. Lauffenburger, and P.~K.
  Sorger.
\newblock Modeling a snap-action, variable-delay switch controlling extrinsic
  cell death.
\newblock {\em PLoS Biol}, 6(12):e299, 2008.

\bibitem{alhamidi2011fukutin}
M.~Alhamidi, E.~K. Buvang, T.~Fagerheim, V.~Brox, S.~Lindal, M.~Van~Ghelue, and
  {\O}.~Nilssen.
\newblock Fukutin-related protein resides in the golgi cisternae of skeletal
  muscle fibres and forms disulfide-linked homodimers via an n-terminal
  interaction.
\newblock {\em PloS one}, 6(8):e22968, 2011.

\bibitem{amberger2015omim}
J.~S. Amberger, C.~A. Bocchini, F.~Schiettecatte, A.~F. Scott, and A.~Hamosh.
\newblock Omim. org: Online mendelian inheritance in man
  (omim{\textregistered}), an online catalog of human genes and genetic
  disorders.
\newblock {\em Nucleic acids research}, 43(D1):D789--D798, 2015.

\bibitem{arnold2007inflammatory}
L.~Arnold, A.~Henry, F.~Poron, Y.~Baba-Amer, N.~Van~Rooijen, A.~Plonquet, R.~K.
  Gherardi, and B.~Chazaud.
\newblock Inflammatory monocytes recruited after skeletal muscle injury switch
  into antiinflammatory macrophages to support myogenesis.
\newblock {\em The Journal of experimental medicine}, 204(5):1057--1069, 2007.

\bibitem{baker2005dominant}
N.~L. Baker, M.~M{\"o}rgelin, R.~Peat, N.~Goemans, K.~N. North, J.~F. Bateman,
  and S.~R. Lamand{\'e}.
\newblock Dominant collagen vi mutations are a common cause of ullrich
  congenital muscular dystrophy.
\newblock {\em Human molecular genetics}, 14(2):279--293, 2005.

\bibitem{bartoli2008mannosidase}
M.~Bartoli, E.~Gicquel, L.~Barrault, T.~Soheili, M.~Malissen, B.~Malissen,
  N.~Vincent-Lacaze, N.~Perez, B.~Udd, O.~Danos, et~al.
\newblock Mannosidase i inhibition rescues the human $\alpha$-sarcoglycan r77c
  recurrent mutation.
\newblock {\em Human molecular genetics}, 17(9):1214--1221, 2008.

\bibitem{bausch1998local}
A.~R. Bausch, F.~Ziemann, A.~A. Boulbitch, K.~Jacobson, and E.~Sackmann.
\newblock Local measurements of viscoelastic parameters of adherent cell
  surfaces by magnetic bead microrheometry.
\newblock {\em Biophysical journal}, 75(4):2038--2049, 1998.

\bibitem{beggs1991exploring}
A.~Beggs, E.~Hoffman, J.~Snyder, K.~Arahata, L.~Specht, F.~Shapiro,
  C.~Angelini, H.~Sugita, and L.~Kunkel.
\newblock Exploring the molecular basis for variability among patients with
  becker muscular dystrophy: dystrophin gene and protein studies.
\newblock {\em American journal of human genetics}, 49(1):54, 1991.

\bibitem{bhattacharjee2012expression}
R.~B. Bhattacharjee, T.~Zannat, and J.~Bag.
\newblock Expression of the polyalanine expansion mutant of nuclear poly
  (a)-binding protein induces apoptosis via the p53 pathway.
\newblock {\em Cell biology international}, 36(8):697--704, 2012.

\bibitem{bogdanovich2008myostatin}
S.~Bogdanovich, E.~M. McNally, and T.~S. Khurana.
\newblock Myostatin blockade improves function but not histopathology in a
  murine model of limb-girdle muscular dystrophy 2c.
\newblock {\em Muscle \& nerve}, 37(3):308--316, 2008.

\bibitem{brockington2001}
M.~Brockington, D.~J. Blake, P.~Prandini, S.~C. Brown, S.~Torelli, M.~A.
  Benson, C.~P. Ponting, B.~Estournet, N.~B. Romero, E.~Mercuri, et~al.
\newblock Mutations in the fukutin-related protein gene (fkrp) cause a form of
  congenital muscular dystrophy with secondary laminin $\alpha$2 deficiency and
  abnormal glycosylation of $\alpha$-dystroglycan.
\newblock {\em The American Journal of Human Genetics}, 69(6):1198--1209, 2001.

\bibitem{brockington2001mutations}
M.~Brockington, Y.~Yuva, P.~Prandini, S.~C. Brown, S.~Torelli, M.~A. Benson,
  R.~Herrmann, L.~V. Anderson, R.~Bashir, J.-M. Burgunder, et~al.
\newblock Mutations in the fukutin-related protein gene (fkrp) identify limb
  girdle muscular dystrophy 2i as a milder allelic variant of congenital
  muscular dystrophy mdc1c.
\newblock {\em Human molecular genetics}, 10(25):2851--2859, 2001.

\bibitem{brooke1989duchenne}
M.~Brooke, G.~Fenichel, R.~Griggs, J.~Mendell, R.~Moxley, J.~Florence, W.~King,
  S.~Pandya, J.~Robison, J.~Schierbecker, et~al.
\newblock Duchenne muscular dystrophy patterns of clinical progression and
  effects of supportive therapy.
\newblock {\em Neurology}, 39(4):475--475, 1989.

\bibitem{burns2015evidence}
D.~P. Burns and K.~D. O’Halloran.
\newblock Evidence of hypoxic tolerance in weak upper airway muscle from young
  mdx mice.
\newblock {\em Respiratory physiology \& neurobiology}, 2015.

\bibitem{cariaso2012snpedia}
M.~Cariaso and G.~Lennon.
\newblock Snpedia: a wiki supporting personal genome annotation, interpretation
  and analysis.
\newblock {\em Nucleic acids research}, 40(D1):D1308--D1312, 2012.

\bibitem{cheng2011collagen}
I.~Cheng, Y.-C. Lin, E.~Hwang, H.-T. Huang, W.-H. Chang, Y.-L. Liu, and C.-Y.
  Chao.
\newblock Collagen vi protects against neuronal apoptosis elicited by
  ultraviolet irradiation via an akt/phosphatidylinositol 3-kinase signaling
  pathway.
\newblock {\em Neuroscience}, 183:178--188, 2011.

\bibitem{childers2002eccentric}
M.~K. Childers, C.~S. Okamura, D.~J. Bogan, J.~R. Bogan, G.~F. Petroski,
  K.~McDonald, and J.~N. Kornegay.
\newblock Eccentric contraction injury in dystrophic canine muscle.
\newblock {\em Archives of physical medicine and rehabilitation},
  83(11):1572--1578, 2002.

\bibitem{10002015global}
.~G.~P. Consortium et~al.
\newblock A global reference for human genetic variation.
\newblock {\em Nature}, 526(7571):68--74, 2015.

\bibitem{Cozzolino201254}
M.~Cozzolino and M.~T. Carrì.
\newblock Mitochondrial dysfunction in \{ALS\}.
\newblock {\em Progress in Neurobiology}, 97(2):54 -- 66, 2012.
\newblock The Neurotoxicity of Mutant Proteins 20 Years after the discovery of
  the first mutant gene involved in neurodegeneration.

\bibitem{de2002mutations}
D.~B.-V. de~Bernab{\'e}, S.~Currier, A.~Steinbrecher, J.~Celli, E.~van
  Beusekom, B.~van~der Zwaag, H.~Kayserili, L.~Merlini, D.~Chitayat, W.~B.
  Dobyns, et~al.
\newblock Mutations in the o-mannosyltransferase gene pomt1 give rise to the
  severe neuronal migration disorder walker-warburg syndrome.
\newblock {\em The American Journal of Human Genetics}, 71(5):1033--1043, 2002.

\bibitem{DellAcqua2009283}
G.~Dell’Acqua and F.~Castiglione.
\newblock Stability and phase transitions in a mathematical model of duchenne
  muscular dystrophy.
\newblock {\em Journal of Theoretical Biology}, 260(2):283 -- 289, 2009.

\bibitem{desguerre2009endomysial}
I.~Desguerre, M.~Mayer, F.~Leturcq, J.-P. Barbet, R.~K. Gherardi, and
  C.~Christov.
\newblock Endomysial fibrosis in duchenne muscular dystrophy: a marker of poor
  outcome associated with macrophage alternative activation.
\newblock {\em Journal of Neuropathology \& Experimental Neurology},
  68(7):762--773, 2009.

\bibitem{dinccer2003novel}
P.~Din{\c{c}}er, B.~Balc{\i}, Y.~Yuva, B.~Talim, M.~Brockington,
  D.~Din{\c{c}}el, S.~Torelli, S.~Brown, G.~Kale, G.~Haliloglu, et~al.
\newblock A novel form of recessive limb girdle muscular dystrophy with mental
  retardation and abnormal expression of $\alpha$-dystroglycan.
\newblock {\em Neuromuscular Disorders}, 13(10):771--778, 2003.

\bibitem{duclos1998progressive}
F.~Duclos, V.~Straub, S.~A. Moore, D.~P. Venzke, R.~F. Hrstka, R.~H. Crosbie,
  M.~Durbeej, C.~S. Lebakken, A.~J. Ettinger, J.~Van Der~Meulen, et~al.
\newblock Progressive muscular dystrophy in $\alpha$-sarcoglycan--deficient
  mice.
\newblock {\em The Journal of cell biology}, 142(6):1461--1471, 1998.

\bibitem{esapa2002functional}
C.~T. Esapa, M.~A. Benson, J.~E. Schr{\"o}der, E.~Martin-Rendon,
  M.~Brockington, S.~C. Brown, F.~Muntoni, S.~Kr{\"o}ger, and D.~J. Blake.
\newblock Functional requirements for fukutin-related protein in the golgi
  apparatus.
\newblock {\em Human molecular genetics}, 11(26):3319--3331, 2002.

\bibitem{fan2001oligomerization}
X.~Fan, P.~Dion, J.~Laganiere, B.~Brais, and G.~A. Rouleau.
\newblock Oligomerization of polyalanine expanded pabpn1 facilitates nuclear
  protein aggregation that is associated with cell death.
\newblock {\em Human molecular genetics}, 10(21):2341--2351, 2001.

\bibitem{friedman2000using}
N.~Friedman, M.~Linial, I.~Nachman, and D.~Pe'er.
\newblock Using bayesian networks to analyze expression data.
\newblock {\em Journal of computational biology}, 7(3-4):601--620, 2000.

\bibitem{fussenegger2000mathematical}
M.~Fussenegger, J.~E. Bailey, and J.~Varner.
\newblock A mathematical model of caspase function in apoptosis.
\newblock {\em Nature biotechnology}, 18(7):768--774, 2000.

\bibitem{gawlik2015deletion}
K.~I. Gawlik and M.~Durbeej.
\newblock Deletion of integrin $\alpha$7 subunit does not aggravate the
  phenotype of laminin $\alpha$2 chain-deficient mice.
\newblock {\em Scientific reports}, 5, 2015.

\bibitem{girgenrath2004inhibition}
M.~Girgenrath, J.~A. Dominov, C.~A. Kostek, and J.~B. Miller.
\newblock Inhibition of apoptosis improves outcome in a model of congenital
  muscular dystrophy.
\newblock {\em The Journal of clinical investigation}, 114(11):1635--1639,
  2004.

\bibitem{glancy2013effect}
B.~Glancy, W.~T. Willis, D.~J. Chess, and R.~S. Balaban.
\newblock Effect of calcium on the oxidative phosphorylation cascade in
  skeletal muscle mitochondria.
\newblock {\em Biochemistry}, 52(16):2793--2809, 2013.

\bibitem{godfrey2007refining}
C.~Godfrey, E.~Clement, R.~Mein, M.~Brockington, J.~Smith, B.~Talim, V.~Straub,
  S.~Robb, R.~Quinlivan, L.~Feng, et~al.
\newblock Refining genotype--phenotype correlations in muscular dystrophies
  with defective glycosylation of dystroglycan.
\newblock {\em Brain}, 130(10):2725--2735, 2007.

\bibitem{grady1999role}
R.~M. Grady, R.~W. Grange, K.~S. Lau, M.~M. Maimone, M.~C. Nichol, J.~T. Stull,
  and J.~R. Sanes.
\newblock Role for $\alpha$-dystrobrevin in the pathogenesis of
  dystrophin-dependent muscular dystrophies.
\newblock {\em Nature Cell Biology}, 1(4):215--220, 1999.

\bibitem{grunwald2008petri}
S.~Grunwald, A.~Speer, J.~Ackermann, and I.~Koch.
\newblock Petri net modelling of gene regulation of the duchenne muscular
  dystrophy.
\newblock {\em Biosystems}, 92(2):189--205, 2008.

\bibitem{Gurpur2009999}
P.~B. Gurpur, J.~Liu, D.~J. Burkin, and S.~J. Kaufman.
\newblock Valproic acid activates the pi3k/akt/mtor pathway in muscle and
  ameliorates pathology in a mouse model of duchenne muscular dystrophy.
\newblock {\em The American Journal of Pathology}, 174(3):999 -- 1008, 2009.

\bibitem{hack1998gamma}
A.~A. Hack, C.~T. Ly, F.~Jiang, C.~J. Clendenin, K.~S. Sigrist, R.~L. Wollmann,
  and E.~M. McNally.
\newblock $\gamma$-sarcoglycan deficiency leads to muscle membrane defects and
  apoptosis independent of dystrophin.
\newblock {\em The Journal of cell biology}, 142(5):1279--1287, 1998.

\bibitem{hall2015guyton}
J.~E. Hall.
\newblock {\em Guyton and Hall textbook of medical physiology}.
\newblock Elsevier Health Sciences, 2015.

\bibitem{heller2015human}
K.~N. Heller, C.~L. Montgomery, K.~M. Shontz, K.~R. Clark, J.~R. Mendell, and
  L.~R. Rodino-Klapac.
\newblock Human $\alpha$7 integrin gene (itga7) delivered by adeno-associated
  virus extends survival of severely affected dystrophin/utrophin-deficient
  mice.
\newblock {\em Human gene therapy}, 26(10):647--656, 2015.

\bibitem{hill1938heat}
A.~Hill.
\newblock The heat of shortening and the dynamic constants of muscle.
\newblock {\em Proceedings of the Royal Society of London B: Biological
  Sciences}, 126(843):136--195, 1938.

\bibitem{hodgson2015modelling}
D.~Hodgson, V.~Mudera, and R.~Torii.
\newblock Modelling muscles in vitro: a finite element analysis study.
\newblock {\em CoMPLEX}, 2015.

\bibitem{hoffman1987dystrophin}
E.~P. Hoffman, R.~H. Brown, and L.~M. Kunkel.
\newblock Dystrophin: the protein product of the duchenne muscular dystrophy
  locus.
\newblock {\em Cell}, 51(6):919--928, 1987.

\bibitem{hoffman1988characterization}
E.~P. Hoffman, K.~H. Fischbeck, R.~H. Brown, M.~Johnson, R.~Medori, J.~D.
  Loire, J.~B. Harris, R.~Waterston, M.~Brooke, L.~Specht, et~al.
\newblock Characterization of dystrophin in muscle-biopsy specimens from
  patients with duchenne's or becker's muscular dystrophy.
\newblock {\em New England Journal of Medicine}, 318(21):1363--1368, 1988.

\bibitem{hoffman1987conservation}
E.~P. Hoffman, A.~P. Monaco, C.~C. Feener, and L.~M. Kunkel.
\newblock Conservation of the duchenne muscular dystrophy gene in mice and
  humans.
\newblock {\em Science}, 238(4825):347--350, 1987.

\bibitem{hoops2006copasi}
S.~Hoops, S.~Sahle, R.~Gauges, C.~Lee, J.~Pahle, N.~Simus, M.~Singhal, L.~Xu,
  P.~Mendes, and U.~Kummer.
\newblock Copasi—a complex pathway simulator.
\newblock {\em Bioinformatics}, 22(24):3067--3074, 2006.

\bibitem{hu2015musculoskeletal}
X.~Hu and S.~S. Blemker.
\newblock Musculoskeletal simulation can help explain selective muscle
  degeneration in duchenne muscular dystrophy.
\newblock {\em Muscle \& nerve}, 52(2):174--182, 2015.

\bibitem{huber2010diffusion}
H.~J. Huber, M.~A. Laussmann, J.~H. Prehn, and M.~Rehm.
\newblock Diffusion is capable of translating anisotropic apoptosis initiation
  into a homogeneous execution of cell death.
\newblock {\em BMC systems biology}, 4(1):1, 2010.

\bibitem{huxley1974muscular}
A.~Huxley.
\newblock Muscular contraction.
\newblock {\em The Journal of physiology}, 243(1):1, 1974.

\bibitem{huxley1957muscle}
A.~F. Huxley.
\newblock Muscle structure and theories of contraction.
\newblock {\em Prog. Biophys. Biophys. Chem}, 7:255--318, 1957.

\bibitem{jarrah2014mathematical}
A.~S. Jarrah, F.~Castiglione, N.~P. Evans, R.~W. Grange, and R.~Laubenbacher.
\newblock A mathematical model of skeletal muscle disease and immune response
  in the mdx mouse.
\newblock {\em BioMed research international}, 2014, 2014.

\bibitem{jewell1958analysis}
B.~Jewell and D.~Wilkie.
\newblock An analysis of the mechanical components in frog's striated muscle.
\newblock {\em The Journal of physiology}, 143(3):515, 1958.

\bibitem{johansson2000finite}
T.~Johansson, P.~Meier, and R.~Blickhan.
\newblock A finite-element model for the mechanical analysis of skeletal
  muscles.
\newblock {\em Journal of Theoretical Biology}, 206(1):131--149, 2000.

\bibitem{judge2006dissecting}
L.~M. Judge, M.~Haraguchiln, and J.~S. Chamberlain.
\newblock Dissecting the signaling and mechanical functions of the
  dystrophin-glycoprotein complex.
\newblock {\em Journal of cell science}, 119(8):1537--1546, 2006.

\bibitem{kim2004pomt1}
D.-S. Kim, Y.~Hayashi, H.~Matsumoto, M.~Ogawa, S.~Noguchi, N.~Murakami,
  R.~Sakuta, M.~Mochizuki, D.~Michele, K.~Campbell, et~al.
\newblock Pomt1 mutation results in defective glycosylation and loss of
  laminin-binding activity in $\alpha$-dg.
\newblock {\em Neurology}, 62(6):1009--1011, 2004.

\bibitem{kowald1994towards}
A.~Kowald and T.~Kirkwood.
\newblock Towards a network theory of ageing: a model combining the free
  radical theory and the protein error theory.
\newblock {\em Journal of theoretical Biology}, 168(1):75--94, 1994.

\bibitem{kowald1996network}
A.~Kowald and T.~Kirkwood.
\newblock A network theory of ageing: the interactions of defective
  mitochondria, aberrant proteins, free radicals and scavengers in the ageing
  process.
\newblock {\em Mutation Research/DNAging}, 316(5):209--236, 1996.

\bibitem{lammerding2004lamin}
J.~Lammerding, P.~C. Schulze, T.~Takahashi, S.~Kozlov, T.~Sullivan, R.~D. Kamm,
  C.~L. Stewart, and R.~T. Lee.
\newblock Lamin a/c deficiency causes defective nuclear mechanics and
  mechanotransduction.
\newblock {\em The Journal of clinical investigation}, 113(3):370--378, 2004.

\bibitem{lehman1994ca2+}
W.~Lehman, R.~Craig, and P.~Vibert.
\newblock Ca2+-induced tropomyosin movement in limulus thin filaments revealed
  by three-dimensional reconstruction.
\newblock 1994.

\bibitem{lemos2015nilotinib}
D.~R. Lemos, F.~Babaeijandaghi, M.~Low, C.-K. Chang, S.~T. Lee, D.~Fiore, R.-H.
  Zhang, A.~Natarajan, S.~A. Nedospasov, and F.~M. Rossi.
\newblock Nilotinib reduces muscle fibrosis in chronic muscle injury by
  promoting tnf-mediated apoptosis of fibro/adipogenic progenitors.
\newblock {\em Nature medicine}, 21(7):786--794, 2015.

\bibitem{li2004yeast}
F.~Li, T.~Long, Y.~Lu, Q.~Ouyang, and C.~Tang.
\newblock The yeast cell-cycle network is robustly designed.
\newblock {\em Proceedings of the National Academy of Sciences of the United
  States of America}, 101(14):4781--4786, 2004.

\bibitem{lin1993structural}
F.~Lin and H.~J. Worman.
\newblock Structural organization of the human gene encoding nuclear lamin a
  and nuclear lamin c.
\newblock {\em Journal of Biological Chemistry}, 268(22):16321--16326, 1993.

\bibitem{long2014prevention}
C.~Long, J.~R. McAnally, J.~M. Shelton, A.~A. Mireault, R.~Bassel-Duby, and
  E.~N. Olson.
\newblock Prevention of muscular dystrophy in mice by crispr/cas9--mediated
  editing of germline dna.
\newblock {\em Science}, 345(6201):1184--1188, 2014.

\bibitem{lucas1995muscular}
B.~Lucas-Heron.
\newblock Muscular degeneration in duchenne's dystrophy may be caused by a
  mitochondrial defect.
\newblock {\em Medical hypotheses}, 44(4):298--300, 1995.

\bibitem{lucas1996skeletal}
B.~Lucas-Heron.
\newblock Skeletal muscle of patients with duchenne's muscular dystrophy:
  Evidence of a mitochondrial proteolytic factor responsible for calmitine
  deficiency.
\newblock {\em Biochemical and biophysical research communications},
  223(1):31--35, 1996.

\bibitem{mendell2012gene}
J.~R. Mendell, L.~Rodino-Klapac, Z.~Sahenk, V.~Malik, B.~K. Kaspar, C.~M.
  Walker, and K.~R. Clark.
\newblock Gene therapy for muscular dystrophy: lessons learned and path
  forward.
\newblock {\em Neuroscience letters}, 527(2):90--99, 2012.

\bibitem{mendell2010sustained}
J.~R. Mendell, L.~R. Rodino-Klapac, X.~Q. Rosales, B.~D. Coley, G.~Galloway,
  S.~Lewis, V.~Malik, C.~Shilling, B.~J. Byrne, T.~Conlon, et~al.
\newblock Sustained alpha-sarcoglycan gene expression after gene transfer in
  limb-girdle muscular dystrophy, type 2d.
\newblock {\em Annals of neurology}, 68(5):629--638, 2010.

\bibitem{mills2000m}
C.~D. Mills, K.~Kincaid, J.~M. Alt, M.~J. Heilman, and A.~M. Hill.
\newblock M-1/m-2 macrophages and the th1/th2 paradigm.
\newblock {\em The Journal of Immunology}, 164(12):6166--6173, 2000.

\bibitem{nachman2004inferring}
I.~Nachman, A.~Regev, and N.~Friedman.
\newblock Inferring quantitative models of regulatory networks from expression
  data.
\newblock {\em Bioinformatics}, 20(suppl 1):i248--i256, 2004.

\bibitem{norwood2009prevalence}
F.~L. Norwood, C.~Harling, P.~F. Chinnery, M.~Eagle, K.~Bushby, and V.~Straub.
\newblock Prevalence of genetic muscle disease in northern england: in-depth
  analysis of a muscle clinic population.
\newblock {\em Brain}, page awp236, 2009.

\bibitem{parsons2006age}
S.~A. Parsons, D.~P. Millay, M.~A. Sargent, E.~M. McNally, and J.~D. Molkentin.
\newblock Age-dependent effect of myostatin blockade on disease severity in a
  murine model of limb-girdle muscular dystrophy.
\newblock {\em The American journal of pathology}, 168(6):1975--1985, 2006.

\bibitem{pate1989model}
E.~Pate and R.~Cooke.
\newblock A model of crossbridge action: the effects of atp, adp and pi.
\newblock {\em Journal of Muscle Research \& Cell Motility}, 10(3):181--196,
  1989.

\bibitem{pauly2012ampk}
M.~Pauly, F.~Daussin, Y.~Burelle, T.~Li, R.~Godin, J.~Fauconnier,
  C.~Koechlin-Ramonatxo, G.~Hugon, A.~Lacampagne, M.~Coisy-Quivy, et~al.
\newblock Ampk activation stimulates autophagy and ameliorates muscular
  dystrophy in the mdx mouse diaphragm.
\newblock {\em The American journal of pathology}, 181(2):583--592, 2012.

\bibitem{Pieczenik200784}
S.~R. Pieczenik and J.~Neustadt.
\newblock Mitochondrial dysfunction and molecular pathways of disease.
\newblock {\em Experimental and Molecular Pathology}, 83(1):84 -- 92, 2007.

\bibitem{pozsgai2014gp}
E.~Pozsgai, D.~Griffin, K.~Heller, J.~Mendell, and L.~Rodino-Klapac.
\newblock Gp 227: Beta-sarcoglycan gene transfer leads to functional
  improvement in a model of lgmd2e.
\newblock {\em Neuromuscular Disorders}, 24(9):885, 2014.

\bibitem{rao1996lamin}
L.~Rao, D.~Perez, and E.~White.
\newblock Lamin proteolysis facilitates nuclear events during apoptosis.
\newblock {\em The Journal of cell biology}, 135(6):1441--1455, 1996.

\bibitem{rehm2009dynamics}
M.~Rehm, H.~J. Huber, C.~T. Hellwig, S.~Anguissola, H.~Dussmann, and J.~H.
  Prehn.
\newblock Dynamics of outer mitochondrial membrane permeabilization during
  apoptosis.
\newblock {\em Cell Death \& Differentiation}, 16(4):613--623, 2009.

\bibitem{rodino2009inhibition}
L.~R. Rodino-Klapac, A.~M. Haidet, J.~Kota, C.~Handy, B.~K. Kaspar, and J.~R.
  Mendell.
\newblock Inhibition of myostatin with emphasis on follistatin as a therapy for
  muscle disease.
\newblock {\em Muscle \& nerve}, 39(3):283--296, 2009.

\bibitem{rohr2010snoopy}
C.~Rohr, W.~Marwan, and M.~Heiner.
\newblock Snoopy—a unifying petri net framework to investigate biomolecular
  networks.
\newblock {\em Bioinformatics}, 26(7):974--975, 2010.

\bibitem{rolls2002targeting}
M.~M. Rolls, D.~H. Hall, M.~Victor, E.~H. Stelzer, and T.~A. Rapoport.
\newblock Targeting of rough endoplasmic reticulum membrane proteins and
  ribosomes in invertebrate neurons.
\newblock {\em Molecular biology of the cell}, 13(5):1778--1791, 2002.

\bibitem{romitti2015prevalence}
P.~A. Romitti, Y.~Zhu, S.~Puzhankara, K.~A. James, S.~K. Nabukera, G.~K. Zamba,
  E.~Ciafaloni, C.~Cunniff, C.~M. Druschel, K.~D. Mathews, et~al.
\newblock Prevalence of duchenne and becker muscular dystrophies in the united
  states.
\newblock {\em Pediatrics}, 135(3):513--521, 2015.

\bibitem{rybakova2000dystrophin}
I.~N. Rybakova, J.~R. Patel, and J.~M. Ervasti.
\newblock The dystrophin complex forms a mechanically strong link between the
  sarcolemma and costameric actin.
\newblock {\em The Journal of cell biology}, 150(5):1209--1214, 2000.

\bibitem{sakaki2001interaction}
M.~Sakaki, H.~Koike, N.~Takahashi, N.~Sasagawa, S.~Tomioka, K.~Arahata, and
  S.~Ishiura.
\newblock Interaction between emerin and nuclear lamins.
\newblock {\em Journal of Biochemistry}, 129(2):321--327, 2001.

\bibitem{salmikangas2003myotilin}
P.~Salmikangas, P.~F. van~der Ven, M.~Lalowski, A.~Taivainen, F.~Zhao,
  H.~Suila, R.~Schr{\"o}der, P.~Lappalainen, D.~O. F{\"u}rst, and
  O.~Carp{\'e}n.
\newblock Myotilin, the limb-girdle muscular dystrophy 1a (lgmd1a) protein,
  cross-links actin filaments and controls sarcomere assembly.
\newblock {\em Human molecular genetics}, 12(2):189--203, 2003.

\bibitem{sewry2001skeletal}
C.~Sewry, S.~Brown, E.~Mercuri, G.~Bonne, L.~Feng, G.~Camici, G.~Morris, and
  F.~Muntoni.
\newblock Skeletal muscle pathology in autosomal dominant emery-dreifuss
  muscular dystrophy with lamin a/c mutations.
\newblock {\em Neuropathology and applied neurobiology}, 27(4):281--290, 2001.

\bibitem{sewry1996abnormalities}
C.~Sewry, J.~Taylor, L.~Anderson, E.~Ozawa, R.~Pogue, F.~Piccolo, K.~Bushby,
  V.~Dubowitz, and F.~Muntoni.
\newblock Abnormalities in $\alpha$-, $\beta$-and $\gamma$-sarcoglycan in
  patients with limb-girdle muscular dystrophy.
\newblock {\em Neuromuscular Disorders}, 6(6):467--474, 1996.

\bibitem{sharafi2010micromechanical}
B.~Sharafi and S.~S. Blemker.
\newblock A micromechanical model of skeletal muscle to explore the effects of
  fiber and fascicle geometry.
\newblock {\em Journal of biomechanics}, 43(16):3207--3213, 2010.

\bibitem{smith2012characterization}
A.~Smith, S.~Passey, L.~Greensmith, V.~Mudera, and M.~Lewis.
\newblock Characterization and optimization of a simple, repeatable system for
  the long term in vitro culture of aligned myotubes in 3d.
\newblock {\em Journal of cellular biochemistry}, 113(3):1044--1053, 2012.

\bibitem{spencer2001helper}
M.~J. Spencer, E.~Montecino-Rodriguez, K.~Dorshkind, and J.~G. Tidball.
\newblock Helper (cd4+) and cytotoxic (cd8+) t cells promote the pathology of
  dystrophin-deficient muscle.
\newblock {\em Clinical immunology}, 98(2):235--243, 2001.

\bibitem{spencer1997myonuclear}
M.~J. Spencer, C.~M. Walsh, K.~A. Dorshkind, E.~M. Rodriguez, and J.~G.
  Tidball.
\newblock Myonuclear apoptosis in dystrophic mdx muscle occurs by
  perforin-mediated cytotoxicity.
\newblock {\em Journal of Clinical Investigation}, 99(11):2745, 1997.

\bibitem{st1994differential}
B.~St~Pierre and J.~G. Tidball.
\newblock Differential response of macrophage subpopulations to soleus muscle
  reloading after rat hindlimb suspension.
\newblock {\em Journal of Applied Physiology}, 77(1):290--297, 1994.

\bibitem{sullivan1999loss}
T.~Sullivan, D.~Escalante-Alcalde, H.~Bhatt, M.~Anver, N.~Bhat, K.~Nagashima,
  C.~L. Stewart, and B.~Burke.
\newblock Loss of a-type lamin expression compromises nuclear envelope
  integrity leading to muscular dystrophy.
\newblock {\em The Journal of cell biology}, 147(5):913--920, 1999.

\bibitem{tam2013mathematical}
Z.~Y. Tam, J.~Gruber, B.~Halliwell, and R.~Gunawan.
\newblock Mathematical modeling of the role of mitochondrial fusion and fission
  in mitochondrial dna maintenance.
\newblock {\em PloS one}, 8(10):e76230, 2013.

\bibitem{ter1998laminin}
H.~Ter~Laak, Q.~Leyten, F.~Gabre{\"e}ls, H.~Kuppen, W.~Renier, and R.~Sengers.
\newblock Laminin-$\alpha$ 2 (merosin), $\beta$-dystroglycan,
  $\alpha$-sarcoglycan (adhalin), and dystrophin expression in congenital
  muscular dystrophies: An immunohistochemical study.
\newblock {\em Clinical neurology and neurosurgery}, 100(1):5--10, 1998.

\bibitem{tidball1995apoptosis}
J.~G. Tidball, D.~E. Albrecht, B.~E. Lokensgard, and M.~J. Spencer.
\newblock Apoptosis precedes necrosis of dystrophin-deficient muscle.
\newblock {\em Journal of cell science}, 108(6):2197--2204, 1995.

\bibitem{trollet2010molecular}
C.~Trollet, S.~Y. Anvar, A.~Venema, I.~P. Hargreaves, K.~Foster, A.~Vignaud,
  A.~Ferry, E.~Negroni, C.~Hourde, M.~A. Baraibar, et~al.
\newblock Molecular and phenotypic characterization of a mouse model of
  oculopharyngeal muscular dystrophy reveals severe muscular atrophy restricted
  to fast glycolytic fibres.
\newblock {\em Human molecular genetics}, page ddq098, 2010.

\bibitem{tsuchiya1999distinct}
Y.~Tsuchiya, A.~Hase, M.~Ogawa, H.~Yorifuji, and K.~Arahata.
\newblock Distinct regions specify the nuclear membrane targeting of emerin,
  the responsible protein for emery--dreifuss muscular dystrophy.
\newblock {\em European Journal of Biochemistry}, 259(3):859--865, 1999.

\bibitem{tucker2006temporal}
A.~Tucker, P.~Hoen, V.~Vinciotti, and X.~Liu.
\newblock Temporal bayesian classifiers for modelling muscular dystrophy
  expression data.
\newblock {\em Intelligent Data Analysis}, 10(5):441--455, 2006.

\bibitem{van1998modelling}
B.~Van~der Linden, H.~Koopman, H.~Grootenboer, and P.~Huijing.
\newblock Modelling functional effects of muscle geometry.
\newblock {\em Journal of Electromyography and kinesiology}, 8(2):101--109,
  1998.

\bibitem{van1998revised}
B.~Van~der Linden, H.~Koopman, P.~Huijing, and H.~Grootenboer.
\newblock Revised planimetric model of unipennate skeletal muscle: a mechanical
  approach.
\newblock {\em Clinical Biomechanics}, 13(4):256--260, 1998.

\bibitem{van1997functional}
J.~Van~Leeuwen and W.~M. Kier.
\newblock Functional design of tentacles in squid: linking sarcomere
  ultrastructure to gross morphological dynamics.
\newblock {\em Philosophical Transactions of the Royal Society of London B:
  Biological Sciences}, 352(1353):551--571, 1997.

\bibitem{virgilio2015multiscale}
K.~M. Virgilio, K.~S. Martin, S.~M. Peirce, and S.~S. Blemker.
\newblock Multiscale models of skeletal muscle reveal the complex effects of
  muscular dystrophy on tissue mechanics and damage susceptibility.
\newblock {\em Interface focus}, 5(2):20140080, 2015.

\bibitem{wehling2001nitric}
M.~Wehling, M.~J. Spencer, and J.~G. Tidball.
\newblock A nitric oxide synthase transgene ameliorates muscular dystrophy in
  mdx mice.
\newblock {\em The Journal of cell biology}, 155(1):123--132, 2001.

\bibitem{williams2006sarcolemmal}
J.~C. Williams, A.~L. Armesilla, T.~M. Mohamed, C.~L. Hagarty, F.~H. McIntyre,
  S.~Schomburg, A.~O. Zaki, D.~Oceandy, E.~J. Cartwright, M.~H. Buch, et~al.
\newblock The sarcolemmal calcium pump, $\alpha$-1 syntrophin, and neuronal
  nitric-oxide synthase are parts of a macromolecular protein complex.
\newblock {\em Journal of Biological Chemistry}, 281(33):23341--23348, 2006.

\bibitem{williams1999differential}
M.~W. Williams and R.~J. Bloch.
\newblock Differential distribution of dystrophin and $\beta$-spectrin at the
  sarcolemma of fast twitch skeletal muscle fibers.
\newblock {\em Journal of Muscle Research \& Cell Motility}, 20(4):383--393,
  1999.

\bibitem{Winklhofer201029}
K.~F. Winklhofer and C.~Haass.
\newblock Mitochondrial dysfunction in parkinson's disease.
\newblock {\em Biochimica et Biophysica Acta (BBA) - Molecular Basis of
  Disease}, 1802(1):29 -- 44, 2010.
\newblock Mitochondrial Dysfunction.

\bibitem{xu2015crispr}
L.~Xu, K.~H. Park, L.~Zhao, J.~Xu, M.~El~Refaey, Y.~Gao, H.~Zhu, J.~Ma, and
  R.~Han.
\newblock Crispr-mediated genome editing restores dystrophin expression and
  function in mdx mice.
\newblock {\em Molecular Therapy}, 2015.

\bibitem{yucesoy2002three}
C.~A. Yucesoy, B.~H. Koopman, P.~A. Huijing, and H.~J. Grootenboer.
\newblock Three-dimensional finite element modeling of skeletal muscle using a
  two-domain approach: linked fiber-matrix mesh model.
\newblock {\em Journal of biomechanics}, 35(9):1253--1262, 2002.

\bibitem{zollner2015high}
A.~M. Z{\"o}llner, J.~M. Pok, E.~J. McWalter, G.~E. Gold, and E.~Kuhl.
\newblock On high heels and short muscles: A multiscale model for sarcomere
  loss in the gastrocnemius muscle.
\newblock {\em Journal of theoretical biology}, 365:301--310, 2015.

\end{thebibliography}

\end{document}